\documentclass[12pt]{article}
\usepackage{graphics,graphicx}
\usepackage{amssymb,epsfig,amsmath,euscript,array}
\usepackage{cite}

\makeatletter
\@addtoreset{equation}{section}
\makeatother

\newcounter{multieqs}

\newcommand{\be}{\begin{equation}}
\newcommand{\ee}{\end{equation}}
\newcommand{\eq}[1]{(\ref{#1})}

\newcommand{\bm}[1]{\mbox{\boldmath $#1$}}

\newcommand{\kslash}{k \!\!\! / }

\newcommand{\lslash}{l \!\! / }
\newcommand{\Pslash}{P \!\!\!\! / }

\newcommand{\islash}{i \!\!\! / }
\newcommand{\jslash}{j \!\!\! / }
\newcommand{\aslash}{a \!\!\! / }
\newcommand{\bslash}{{b \hspace{-6pt} \slash} }

\newcommand{\onslash}{1 \!\!\! / }
\newcommand{\twslash}{2 \!\!\!/ }
\newcommand{\thslash}{3 \!\!\!/ }
\newcommand{\foslash}{4 \!\!\! / }
\newcommand{\fislash}{5 \!\!\! / }

\newcommand{\mslash}{m \!\!\! / }

\def\bd{\begin{document}}
\def\ed{\end{document}}
\def\nn{\nonumber}
\def\bea{\begin{eqnarray}}
\def\eea{\end{eqnarray}}

\def\ab{(ijab)}
\def\ba{(ijba)}
\def\ijab{{\tr}_{+}(\islash\, \jslash\, \aslash \, \bslash)}
\def\ijba{{\tr}_{+}(\islash\, \jslash\, \bslash \, \aslash)}
\def\ijaP{{\tr}_{+}(\islash\, \jslash\, \aslash \, \Pslash)}
\def\ijPLa{{\tr}_{+}(\islash\, \jslash\, \Pslash_L \, \aslash)}
\def\ijaPL{{\tr}_{+}(\islash\, \jslash\, \aslash \, \Pslash_L)}
\def\ijPLza{{\tr}_{+}(\islash\, \jslash\, \Pslash_{L;z} \, \aslash)}
\def\ijaPLz{{\tr}_{+}(\islash\, \jslash\, \aslash \, \Pslash_{L;z})}
\def\ijPa{{\tr}_{+}(\islash\, \jslash\, \Pslash \, \aslash)}
\def\iaPb{{\tr}_{+}(\islash\, \aslash\, \Pslash \, \bslash)}
\def\ibPa{{\tr}_{+}(\islash\, \bslash\, \Pslash \, \aslash)}
\def\ijPmu{{\tr}_{+}(\islash\, \jslash\, \Pslash \, \mu)}
\def\ibmuP{{\tr}_{+}(\islash\, \bslash\, \mu \, \Pslash)}
\def\ibmua{{\tr}_{+}(\islash\, \bslash\, \mu \, \aslash)}
\def\iamub{{\tr}_{+}(\islash\, \aslash\, \mu \, \bslash)}
\def\jaPb{{\tr}_{+}(\jslash\, \aslash\, \Pslash \, \bslash)}
\def\ijmuP{{\tr}_{+}(\islash\, \jslash\, \mu \, \Pslash)}
\def\ijmum{{\tr}_{+}(\islash\, \jslash\, \mu \, \mslash)}
\def\ijmmu{{\tr}_{+}(\islash\, \jslash\, \mslash \, \mu)}
\def\ijmP{{\tr}_{+}(\islash\, \jslash\, \mslash \, \Pslash)}
\def\iabP{{\tr}_{+}(\islash\, \aslash\, \bslash \, \Pslash)}
\def\ijbP{{\tr}_{+}(\islash\, \jslash\, \bslash \, \Pslash)}
\def\jbPa{{\tr}_{+}(\jslash\, \bslash\, \Pslash \, \aslash)}
\def\ijPb{{\tr}_{+}(\islash\, \jslash\, \Pslash \, \bslash)}
\def\jbmua{{\tr}_{+}(\jslash\, \bslash\, \mu \, \aslash)}

\def\loablt{ {\tr}_{+}(\lslash_1\, \aslash \, \bslash\, \lslash_2)}

\def\ijlolt{{\tr}_{+}(\islash\, \jslash\, \lslash_1 \, \lslash_2)}
\def\ijltlo{{\tr}_{+}(\islash\, \jslash\, \lslash_2 \, \lslash_1)}
\def\ibloa{{\tr}_{+}(\islash\, \bslash\, \lslash_1 \, \aslash)}
\def\jaltb{{\tr}_{+}(\jslash\, \aslash\, \lslash_2 \, \bslash)}
\def\ialtb{{\tr}_{+}(\islash\, \aslash\, \lslash_2 \, \bslash)}
\def\bltloa{{\tr}_{+}(\bslash\, \lslash_2\, \lslash_1 \, \aslash)}
\def\jbloa{{\tr}_{+}(\jslash\, \bslash\, \lslash_1 \, \aslash)}
\def\ibPb{{\tr}_{+}(\islash\, \bslash\, \Pslash \, \bslash)}
\def\ijltb{{\tr}_{+}(\islash\, \jslash\, \lslash_2 \, \bslash)}

\def\ijloa{{\tr}_{+}(\islash\, \jslash\,  \lslash_1 \, \aslash)}
\def\ijblt{{\tr}_{+}(\islash\, \jslash\,  \bslash \, \lslash_2)}

\def\jakb{{\tr}_{+}(\jslash\, \aslash\, \kslash \, \bslash)}
\def\iakb{{\tr}_{+}(\islash\, \aslash\, \kslash \, \bslash)}

\def\tofo{{\tr}_{+}(\onslash\, \thslash\, \twslash \, \foslash)}
\def\foto{{\tr}_{+}(\onslash\, \thslash\, \foslash \, \twslash)}
\def\tofi{{\tr}_{+}(\onslash\, \thslash\, \twslash \, \fislash)}
\def\fito{{\tr}_{+}(\onslash\, \thslash\, \fislash \, \twslash)}

\def\lrangle#1#2{\langle #1\,#2\rangle}

\def\Li{{$\rm Li}_2$}
\def\eps{\epsilon}
\def\epsuv{{\epsilon_{\rm \mbox{\tiny UV}}}}
\let\bm=\bibitem
\let\la=\label

\def\npb#1#2#3{Nucl. Phys. {\bf{B#1}} #3 (#2)}
\def\plb#1#2#3{Phys. Lett. {\bf{#1B}} #3 (#2)}
\def\prl#1#2#3{Phys. Rev. Lett. {\bf{#1}} #3 (#2)}
\def\prd#1#2#3{Phys. Rev. {D \bf{#1}} #3 (#2)}
\def\cmp#1#2#3{Comm. Math. Phys. {\bf{#1}} #3 (#2)}
\def\cqg#1#2#3{Class. Quantum Grav. {\bf{#1}} #3 (#2)}
\def\nppsa#1#2#3{Nucl. Phys. B (Proc. Suppl.) {\bf{#1A}}#3 (#2)}
\def\ap#1#2#3{Ann. of Phys. {\bf{#1}} #3 (#2)}
\def\ijmp#1#2#3{Int. J. Mod. Phys. {\bf{A#1}} #3 (#2)}
\def\rmp#1#2#3{Rev. Mod. Phys. {\bf{#1}} #3 (#2)}
\def\mpla#1#2#3{Mod. Phys. Lett. {\bf A#1} #3 (#2)}
\def\jhep#1#2#3{J. High Energy Phys. {\bf #1} #3 (#2)}
\def\atmp#1#2#3{Adv. Theor. Math. Phys. {\bf #1} #3 (#2)}
\newcommand{\EQ}[1]{\begin{equation} #1 \end{equation}}
\newcommand{\AL}[1]{\begin{subequations}\begin{align} #1 \end{align}\end{subequations}}
\newcommand{\SP}[1]{\begin{equation}\begin{split} #1 \end{split}\end{equation}}
\newcommand{\ALAT}[2]{\begin{subequations}\begin{alignat}{#1} #2 \end{alignat}
                        \end{subequations}}
\def\beqa{\begin{eqnarray}}
\def\eeqa{\end{eqnarray}}
\def\beq{\begin{equation}}
\def\eeq{\end{equation}}
\def\sst{\scriptscriptstyle}
\def\thetabar{\bar\theta}
\def\Tr{{\rm Tr}}
\def\one{\mbox{1 \kern-.59em {\rm l}}}
 \def\Nh{\hat{N}}

\newcommand{\half}{{\textstyle {1 \over 2}}}

\def\a{\alpha}      \def\da{{\dot\alpha}}
\def\b{\beta}       \def\db{{\dot\beta}}
\def\c{\gamma}  \def\G{\Gamma}  \def\cdt{\dot\gamma}
\def\d{\delta}  \def\D{\Delta}  \def\ddt{\dot\delta}
\def\e{\epsilon}        \def\vare{\varepsilon}
\def\f{\phi}    \def\F{\Phi}    \def\vvf{\f}
\def\h{\eta}
\def\k{\kappa}
\def\l{\lambda} \def\L{\Lambda}
\def\m{\mu} \def\n{\nu}
\def\o{\omega}
\def\p{\pi} \def\P{\Pi}
\def\r{\rho}
\def\s{\sigma}  \def\S{\Sigma}
\def\t{\tau}
\def\th{\theta} \def\Th{\Theta} \def\vth{\vartheta}
\def\X{\Xeta}
\def\z{\zeta}
\def\de{\partial}

\def\cA{{\cal A}} \def\cB{{\cal B}} \def\cC{{\cal C}}
\def\cD{{\cal D}} \def\cE{{\cal E}} \def\cF{{\cal F}}
\def\cG{{\cal G}} \def\cH{{\cal H}} \def\cI{{\cal I}}
\def\cJ{{\cal J}} \def\cK{{\cal K}} \def\cL{{\cal L}}
\def\cM{{\cal M}} \def\cN{{\cal N}} \def\cO{{\cal O}}
\def\cP{{\cal P}} \def\cQ{{\cal Q}} \def\cR{{\cal R}}
\def\cS{{\cal S}} \def\cT{{\cal T}} \def\cU{{\cal U}}
\def\cV{{\cal V}} \def\cW{{\cal W}} \def\cX{{\cal X}}
\def\cY{{\cal Y}} \def\cZ{{\cal Z}}

\def\ua{\underline{\alpha}}
\def\ub{\underline{\phantom{\alpha}}\!\!\!\beta}
\def\uc{\underline{\phantom{\alpha}}\!\!\!\gamma}
\def\um{\underline{\mu}}
\def\ud{\underline\delta}
\def\ue{\underline\epsilon}
\def\una{\underline a}\def\unA{\underline A}
\def\unb{\underline b}\def\unB{\underline B}
\def\unc{\underline c}\def\unC{\underline C}
\def\und{\underline d}\def\unD{\underline D}
\def\une{\underline e}\def\unE{\underline E}
\def\unf{\underline{\phantom{e}}\!\!\!\! f}\def\unF{\underline F}
\def\unm{\underline m}\def\unM{\underline M}
\def\unn{\underline n}\def\unN{\underline N}
\def\unp{\underline{\phantom{a}}\!\!\! p}\def\unP{\underline P}
\def\unq{\underline{\phantom{a}}\!\!\! q}
\def\unQ{\underline{\phantom{A}}\!\!\!\! Q}
\def\unH{\underline{H}}

\def\As {{A \hspace{-6.4pt} \slash}\;}
\def\bs {{b \hspace{-6.4pt} \slash}\;}
\def\Ds {{D \hspace{-6.4pt} \slash}\;}
\def\ds {{\del \hspace{-6.4pt} \slash}\;}
\def\ss {{\s \hspace{-6.4pt} \slash}\;}
\def\ks {{ k \hspace{-6.4pt} \slash}\;}
\def\ps {{p \hspace{-6.4pt} \slash}\;}
\def\pas {{{p_1} \hspace{-6.4pt} \slash}\;}
\def\pbs {{{p_2} \hspace{-6.4pt} \slash}\;}
\def\Ps {{P \hspace{-6.4pt} \slash}\;}
\def\Qs {{Q \hspace{-6.4pt} \slash}\;}

\def\Fh{\hat{F}}
\def\Vh{\hat{V}}
\def\Xh{\hat{X}}
\def\ah{\hat{a}}
\def\xh{\hat{x}}
\def\yh{\hat{y}}
\def\ph{\hat{p}}
\def\xih{\hat{\xi}}
\def\psit{\tilde{\psi}}
\def\Psit{\tilde{\Psi}}
\def\tht{\tilde{\th}}
\def\lt{\tilde{\lambda}}
\def\llt{\tilde{l}}
\def\At{\tilde{A}}
\def\Qt{\tilde{Q}}
\def\Rt{\tilde{R}}
\def\Nt{\tilde{N}}

\def\at{\tilde{a}}
\def\st{\tilde{s}}
\def\ft{\tilde{f}}
\def\pt{\tilde{p}}
\def\qt{\tilde{q}}
\def\vt{\tilde{v}}
\def\nt{\tilde{n}}

\def\delb{\bar{\partial}}
\def\bz{\bar{z}}
\def\bD{\bar{D}}
\def\bB{\bar{B}}

\def\bk{{\bf k}}
\def\bl{{\bf l}}
\def\bp{{\bf p}}
\def\bq{{\bf q}}
\def\br{{\bf r}}
\def\bx{{\bf x}}
\def\by{{\bf y}}
\def\bR{{\bf R}}
\def\bV{{\bf V}}

\def\d{\delta}\def\D{\Delta}\def\ddt{\dot\delta}
\def\pa{\partial} \def\del{\partial}
\def\xx{\times}
\def\uno{\mbox{1 \kern-.59em {\rm l}}}
\def\trp{^{\top}}
\def\inv{^{-1}}
\def\dag{{^{\dagger}}}
\def\pr{^{\prime}}
\def\lan{\langle}
\def\ran{\rangle}
\def\rar{\rightarrow}
\def\lar{\leftarrow}
\def\lrar{\leftrightarrow}
\newcommand{\0}{\,\!}      
\def\one{1\!\!1\,\,}
\def\im{\imath}
\def\jm{\jmath}
\newcommand{\tr}{\mbox{tr}}
\newcommand{\slsh}[1]{/ \!\!\!\! #1}
\def\vac{|0\rangle}
\def\lvac{\langle 0|}
\def\hlf{\frac{1}{2}}
\def\ove#1{\frac{1}{#1}}
\def\Box{\square}
\def\ZZ{\mathbb{Z}}
\def\CC#1{({\bf #1})}
\def\bcomment#1{}
\def\bfhat#1{{\bf \hat{#1}}}
\def\VEV#1{\left\langle #1\right\rangle}
\newcommand{\ex}[1]{{\rm e}^{#1}} \def\ii{{\rm i}}
\def\rr{{\rm r}} \def\rs{{\rm s}}\def\rv{{\rm v}}
\def\ri{{\rm i}}\def\rj{{\rm j}}
\newcommand{\lrbrk}[1]{\left(#1\right)}
\newcommand{\sfrac}[2]{{\textstyle\frac{#1}{#2}}}

\def\Li{{\rm Li}_2}

\font\mybb=msbm10 at 12pt
\def\bb#1{\hbox{\mybb#1}}

\font\myBB=msbm10 at 18pt
\def\BB#1{\hbox{\myBB#1}}

\begin{document}

\begin{flushright}
QMUL-PH-08-11
\end{flushright}

\vspace{20pt}

\begin{center}

{\Large \bf Four-point  Amplitudes in $\mathcal{N} = 8$ Supergravity     }
\\
\vspace{0.3cm}
{\Large \bf and Wilson Loops }
\vspace{11pt}
\vspace{32pt}

{\mbox {\bf A.~Brandhuber, P.~Heslop, A.~Nasti, B.~Spence   and G.~Travaglini}}%
\footnote{
{\sffamily \{\tt a.brandhuber, p.j.heslop, a.nasti, w.j.spence,
g.travaglini\}@qmul.ac.uk }}

{\em Centre for Research in String Theory\\
Department of Physics\\
Queen Mary, University of London\\
Mile End Road, London, E1 4NS\\
United Kingdom
 }

\vspace{30pt} {\bf Abstract}

\end{center}

\noindent
Prompted by recent progress in the study of $\cN=4$  super Yang-Mills
amplitudes, and evidence that similar approaches are relevant to $\cN=8$
supergravity, we investigate possible iterative structures and
applications of Wilson loop techniques in maximal supergravity.
We first consider the two-loop, four-point  MHV scattering
amplitude in $\cN=8$ supergravity, confirming that the infrared divergent parts 
exponentiate, and we give the explicit expression which represents the failure for 
this to occur for the finite part. 
We  observe that each term in the expansion 
of the one- and two-loop amplitudes in the dimensional regularisation parameter $\epsilon$ 
has a uniform degree of transcendentality. 
We then turn to consider Wilson loops in supergravity, 
showing that a natural definition of the loop, involving the Christoffel connection, 
fails to reproduce the one-loop amplitude. An alternative expression, which involves 
the metric explicitly, is shown to have a close relationship with the physical amplitude. 
We find that in a gauge in which the cusp diagrams vanish, the remaining
diagrams for this Wilson loop correctly generate the full one-loop, 
four-point $\cN=8$ supergravity amplitude.

\noindent

\setcounter{page}{0}
\thispagestyle{empty}
\newpage


\section{Introduction}
\setcounter{footnote}{0}

Evidence of recursive structures in the S-matrix of gauge theories has emerged in the past few years.
In 2003 Anastasiou, Bern, Dixon and Kosower (ABDK) \cite{abdk} made the remarkable observation that
the planar, four-point MHV scattering amplitude in \mbox{$\cN\!=\!4$} supersymmetric Yang-Mills theory at two loops can be written
as a polynomial of the one-loop amplitude, plus a kinematic-independent numerical constant.
Subsequently, innovations prompted by twistor string theory, and in particular improved generalised unitarity techniques, have made
possible the calculation of  higher-loop amplitudes both in Yang-Mills and in gravity.
Bern, Dixon and Smirnov (BDS) \cite{bds}  were able to show that the iterative structure  uncovered in \cite{abdk} holds up to three loops, and put forward  a conjecture for the all-loop,
$n$-point MHV amplitude in $\cN=4$ super Yang-Mills at the planar level, in which
the all-loop amplitude is obtained by a suitable exponential of
the one-loop amplitude multiplied by the cusp anomalous dimension.

Iterative structures were first discovered by analysing the soft and collinear behaviour
of amplitudes in gauge theory \cite{ir1,ir2,ir3,ir4,ir5,ir6,ir7,ir8}; the remarkable fact uncovered in
\cite{abdk,bds} was that the finite parts of  the MHV amplitude also follow the same pattern induced
by the expected exponentiation of the  infrared divergences.
The BDS proposal was checked at three loops
in  the four-point case in \cite{bds}, and subsequently in  \cite{2l5pt}
for the two-loop, five-point amplitude.
In a very recent paper \cite{seven}, a discrepancy was found between the
form of the amplitude conjectured by BDS and an explicit two-loop calculation of the six-point amplitude.
The result at six points shows that the structure is that of a polynomial in
the one-loop amplitude,  plus a kinematic-dependent finite remainder function.

In a related development \cite{am},  Alday and Maldacena
proved the correctness of the BDS proposal for the four-point amplitude
at strong  coupling using the AdS/CFT correspondence.%
\footnote{For a recent review, see \cite{fernando}.}
 In their calculation,
the exponentiation of the one-loop amplitude occurs through a
saddle point approximation of the string path integral,
which in the AdS case turns out to be exact.
Furthermore, they showed that
the computation of amplitudes at strong coupling is dual to 
the problem of finding the area of a string ending on a lightlike polygonal loop embedded in the
boundary of AdS space. This,  in turn,  is equivalent to the
method for computing a lightlike polygonal Wilson loop at strong coupling
using the AdS/CFT correspondence, where
the edges of the polygon are determined by the momenta of the scattered particles.
In a subsequent paper  \cite{am2} the same authors showed that the BDS conjecture
should be violated for a sufficiently large number of scattered particles. Further evidence
of a breakdown of the BDS conjecture was also found in \cite{lipa}.

The work of \cite{am} suggested that the calculation of a Wilson loop with the same polygonal
contour could be related  to that of the MHV scattering amplitude even at weak coupling.
This was proved  in  \cite{dks} for the one-loop four-point $\cN=4$ amplitude,
and by three of us in \cite{bht}  for the infinite sequence of  one-loop MHV amplitudes in $\cN=4$
super Yang-Mills.
This surprising Wilson loops/amplitudes  duality was later confirmed
at two loops for the four- \cite{dhks4}, five-\cite{dhks5},
and six-point case \cite{dhksbum,dhks6}.
On the Wilson loop side,  exponentiation naturally emerges  as  a result of the
maximal non-Abelian exponentiation theorem \cite{gatheral,taylor}.
Furthermore, the form of the  four- and five-point expression of the Wilson loop
is determined (up to a constant) by  an anomalous dual conformal Ward identity \cite{dhks5}, and found to be of the form predicted by the BDS ansatz. A similar dual conformal symmetry was found for the integral functions appearing in the
expression of the multi-loop amplitudes in \cite{magic}. Since conformal invariance is not restrictive
enough to fully constrain the $n$-side polygonal Wilson loop for $n\geq 6$, it was perhaps not surprising that at  precisely six points the BDS conjecture turned out to be incorrect \cite{seven}.
It is intriguing however that  the Wilson loops/amplitudes duality does not seem to break down -- 
indeed, the  results of \cite{seven} and \cite{dhks6}  show numerical agreement between 
the Wilson loop and  the six-point gluon amplitude at two loops.

These iterative structures in gauge theory and string theory
have been found at the planar level.
Planarity appears to be a key ingredient of the story -- for instance, the non-planar parts
of the four-point MHV amplitude at two loops do not respect the same iterative structure as the planar
part \cite{abdk}. Planarity would also appear to be an important ingredient in any
relation to integrability -- the cusp anomalous dimension appearing in the BDS proposal
is also determined by an integral equation derived in \cite{bes} using integrability. An analytical
solution to this equation was recently presented in   \cite{bkk}.

It is natural to ask if gravity shares any of these remarkable properties.
Gravity is a non-planar theory, hence it is perhaps even more unexpected to find regularities in the
higher-loop structure of its S-matrix.  However,  the mounting
evidence of interconnections between the maximally supersymmetric theories of $\cN=4$ Yang-Mills and $\cN=8$ supergravity%
\footnote{The paper \cite{living-zvi} reviews the subject up to 2002.}
gives reason to be more optimistic.
Perhaps the potentially most impressive similarity between these two theories is the conjecture that the  $\cN=8$ theory could be ultraviolet finite
\cite{zero,Green:2006gt,bsgf1,Green:2006yu,bsgf2,bsgf3,chalmers},
just like its non-gravitational  maximally supersymmetric cousin.
Furthermore, gravity is also well understood in the infrared thanks to 
the results of \cite{weinberg}, where it was found that infrared singularities can be 
resummed to  the exponential of the one-loop infrared divergences, 
in complete similarity to those of QED  \cite{blochnord,yfs}.

With these motivations in mind, in this paper we would like to initiate a twofold investigation in
$\cN=8$ supergravity. Our first goal will consist in  looking for possible iterative structures and
cross-order relations  using the known  results at one and two loops for the MHV four-point scattering amplitudes.
We confirm that the infrared-divergent parts 
exponentiate, but we observe a failure for 
this to occur for the finite parts, in contradistinction with the four- and five-point amplitudes in
$\cN=4$ Yang-Mills.  
On the other hand, we find  that, similarly to  the $\cN=4$ MHV amplitude, each term  
in the expansion of the one- and two-loop $\cN=8$ MHV amplitudes in the dimensional regularisation parameter $\epsilon$ has a uniform degree of transcendentality (or polylogarithmic weight). This is very intriguing, and leads to the speculation that  maximal transcendentality \cite{klov} 
could be yet another common feature of $\cN=4$ super Yang-Mills and $\cN=8$ supergravity.

Our second aim  is to investigate possible relationships between gravitational scattering amplitudes and gravitational Wilson loops.  This second objective is further motivated by some calculations of
gravity amplitudes in the eikonal approximation \cite{ko,fpvv}, and by our belief that there should
exist a strong link between the eikonal approximation \cite{eik1,eik2,eik3} (performed in specific kinematic regions) and the more recent polygonal Wilson loop calculations (performed without reference to any specific kinematic region).

The rest of the paper is organised as follows.
In the next section we will describe the known one- and two-loop MHV amplitudes in $\cN=8$ supergravity, and use them to show that the two-loop amplitude, minus one half of the square of the one loop amplitude, is finite, consistently with general arguments concerning the exponentiation of infrared divergences in gravity. We give the explicit expression for this finite term. 

In Section 3 we turn to a one-loop Wilson loop calculation.
One candidate for the Wilson loop expression, given by an integral of an exponential involving the Christoffel connection, is shown not to give the one-loop supergravity amplitude correctly. 
A second expression for the gravity Wilson loop is then studied, motivated by its application in the eikonal approximation to gravity. This involves the metric explicitly and is not gauge invariant, however
the failure of gauge invariance is restricted to terms localised at the cusps of the Wilson loop.

The individual cusp diagrams and finite diagrams have the structure expected for 
the $\cN=8$ MHV amplitude (with the tree-level amplitude stripped off);  
however, after summing over all diagrams, we find an incorrect  relative factor  of $-2$
between the infrared-singular and the finite terms in comparison  
to the gravity amplitude. 
This is presumably related to the lack of gauge invariance 
of the Wilson loop at the cusps. 
Motivated by these results, we then turn in Section 4  to consider a gauge where the cusp diagrams vanish, which we call the conformal gauge. 
We show that in this gauge the Wilson loop diagrams, where the propagator connects two non-adjacent segments, precisely yield the full four-point $\cN=8$ supergravity amplitude, including finite and
divergent terms, to all orders in the dimensional regularisation parameter $\epsilon$.
This is in complete analogy to what happens in $\cN=4$ Yang-Mills in
a similar gauge, as we show in Appendices \ref{app-A} and \ref{app-B}.

{\bf Note added:} After this work was completed, the preprint \cite{Naculich:2008ew} appeared,
which overlaps with Section \ref{sec:2loops} of this paper.


\section{MHV amplitudes in $\mathcal{N}=8$ supergravity and \\ iterative structures }
\label{sec:2loops}

In this section we start by briefly reviewing  the expressions of the four-point
MHV amplitude in $\cN\!=\!8$ supergravity at one and two loops, 
and  we then move on to study 
iterative  structures at two loops.

\subsection{Background}
The form of the four-point MHV amplitude at $L$  loops in
maximal supergravity  is very  simple.
It  is  given by the tree-level four-point MHV amplitude  $\cM^{\rm tree}_4 $,
times a scalar (helicity-blind)  function,
\beq\label{fullampl}
\cA_{4}^{(L)}  \ =  \cM^{\rm tree}_4\, \cM_4^{(L)}.
\eeq
This  amplitude was first calculated at one loop in  \cite{gsb} from the $\alpha^\prime \to 0$ limit of
a string theory calculation, and later rederived in  \cite{dn} using string-inspired techniques \cite{bk},
as well  as unitarity \cite{Bern:zx,Bern:1994cg}. The  infinite sequence of one-loop
MHV  amplitudes was obtained  in \cite{bdpr}.
Recently, the four- and five-point MHV amplitudes were also rederived in \cite{nt}
using  MHV diagrams. The two- and three-loop expressions were derived in
\cite{hep-th/9802162}, \cite{bsgf2}, respectively.


At  one loop, the function $\cM_4^{(1)}$  is  simply given by  a sum
of three zero-mass box functions,
\beq
\label{4pt1loop}
\cM_{4}^{(1)}  \ = \   -i \, s\,  t\, u\, \Big( {\kappa \over 2}\Big)^2
\Big[ \cI_{4}^{(1)} (s, t) \, + \cI_{4}^{(1)}(s, u) \, + \, \cI_{4}^{(1)} (u, t) \Big] \ ,
\eeq
where
\beq
\cI_{4}^{(1)} (s, t) \ := \ \int\!\!{d^D l \over (2 \pi)^D} \, {1 \over l^2 (l- p_1)^2 ( l - p_1 - p_2)^2 
(l + p_4)^2 } \
\eeq
is a zero-mass box function with external, cyclically ordered null momenta
$p_1$, $p_2$, $p_3$ and $p_4$, which sum to zero.
We  set $s:= (p_1 + p_2)^2$, $t:= (p_2 + p_3)^2$, $u:= (p_1 + p_3)^2= -s-t$, and $D = 4 - 2 \eps$. 
%
Explicitly 
\begin{align}
\label{ciccio}
  \cI_{4}^{(1)} (s, t) \ &= i\, {c_{\Gamma}\over st} \left[ {2 \over
    \epsilon^2}\left[ (-s)^{-\epsilon}+(-t)^{-\epsilon} \right] -
  \left(\log^2{s\over t}+\pi^2\right)\right]\ ,
\end{align}
where $c_\Gamma\ :={ (4 \pi)^{\epsilon-2}}
{\Gamma(1+\epsilon)\Gamma^2(1-\epsilon) / \Gamma(1-2\epsilon)}$.
Using \eqref{ciccio}, we can rewrite \eqref{4pt1loop} as 
\beq
\label{abm}
\cM_{4}^{(1)}  \ = \ \Big( {\kappa \over 2} \Big)^2 c_\G 
\left[ {2 \over
    \epsilon^2}\left[ (-s)^{1-\epsilon}+(-t)^{1-\epsilon}  + (-u)^{1-\eps} \right] -
  u \log^2{s\over t} - s \log^2{t\over u} - t \log^2{u\over s}
  \right]\ .
  \eeq


The simplicity of \eqref{fullampl}, where the tree-level amplitude factors out  leaving
a helicity-blind function of the particle momenta is clearly reminiscent of
the structure for the infinite sequence of MHV scattering amplitudes in maximally supersymmetric
Yang-Mills. This motivates the search for
\begin{itemize}
\item[{\bf a.}] an iterative structure in the higher-loop amplitude similar to that discovered in
\cite{abdk,bds} for the $\cN=4$ amplitude, and
\item[{\bf b.}]
a  derivation of the functions $\cM_4^{(L)}$ using Wilson loops.
\end{itemize}
The investigation of possible iterative structures of  MHV amplitudes  in  $\cN=4$ Yang-Mills was
motivated by the known structure of the infrared divergences.
This led BDS to propose in \cite{bds} the following conjecture for the all-loop 
MHV amplitude in $\cN=4$ super Yang-Mills: 
\beq
\label{bds}
\cM_{n,\mathrm{YM}} \ := \ 1 + \sum_{L=1}^{\infty} a^L \cM_{n,\mathrm{YM}}^{(L)} (\epsilon )  \ =  \ 
\exp \Big[ \sum_{L=1}^{\infty} a^L  \Big( f^{(L)} (\epsilon) \cM_{n,\mathrm{YM}}^{(1)} ( L \epsilon )  + C^{(L)} + E_n^{(L)}(\epsilon )\Big) 
\Big] 
\ , 
\eeq
where $a=[{g^2 N/ (8 \pi^2)}] (4\pi e^{-\gamma})^\eps$.  In \eqref{bds}, 
$f^{(L)}(\epsilon ) = f_0^{(L)} + f_1^{(L)} \epsilon + f_2^{(L)} \epsilon^2$ is a set of functions, 
one at each loop order, which make their appearance in the exponentiated all-loop expression 
for the  infrared divergences in generic amplitudes in dimensional regularisation \cite{ir6}. 
Specifically, $f_0^{(L)} = \gamma_{K}^{(L)} / 4$, where $\gamma_{K}$ is the cusp anomalous dimension
(related to the anomalous dimension of twist-two operators of large spin). 
Importantly, the  constants $C^{(L)}$ do not depend on the
kinematics or on the number of particles $n$. The non-iterating contributions 
$E_n^{(L)}$ vanish as $\epsilon \to 0$ and depend explicitly on $n$ and the kinematics. 

As we have mentioned in the Introduction, the BDS proposal  has been confirmed 
at two \cite{abdk} and three loops \cite{bds} in the four-point case,  at two loops 
for  the five-point amplitude \cite{2l5pt}, but the recent work of \cite{seven} shows 
a breakdown at two loops in the six-point case \cite{seven}. 
The infrared-singular part of $\cM_{n,\mathrm{YM}}$ is of course correctly reproduced 
by the infrared-divergent part of  the right hand side of \eqref{bds}. 

In order to check \eqref{bds}, one takes the log and expands both sides in perturbation theory; for example, at two loops, one gets
\beq
\label{fam}
\cM_{n,\mathrm{YM}}^{(2)} (\epsilon ) \ - \  {1\over 2} \Big(
\cM_{n,\mathrm{YM}}^{(1)}  (\epsilon ) \Big)^2 \ = \ 
  f^{(2)} (\epsilon) \cM_{n,\mathrm{YM}}^{(1)} ( 2 \epsilon )  \, + \,   C^{(2)}\, +\, E_n^{(2)}
\  . 
\eeq
We wish to follow the same path here for  $\cN=8$ supergravity, starting from the
observation that in gravity the one-loop infrared divergences exponentiate \cite{weinberg}. 
In the four-point case, the leading infrared divergences are expected to resum to 
\beq
\label{expdiv}
\exp \left[ { c_{\Gamma} } \left(   {\kappa \over 2}\right)^2 {2 \over \eps} \, 
\Big( s \, \log(-s) + t \, \log(-t)+u \, \log(-u) \Big) \right] \ .
\eeq
Notice the appearance of the invariant $u = (p_1 + p_3)^2$, due to the lack of colour ordering. 
Moreover, in \cite{bdpr} it was shown that the tree-level soft and collinear splitting  amplitudes in  gravity 
are exact to all orders in perturbation theory. This is due to  the fact that the coupling constant 
$\kappa$ is dimensionful, and it is always accompanied by a power of a kinematic invariant which 
vanishes in the limit considered \cite{weinberg, bdpr}.

We write the four-point MHV amplitude in $\cN=8$ supergravity (stripped of the tree-level prefactor) 
as%
\footnote{Notice  that in \eqref{sotto} we absorb the appropriate power of $\kappa$ in the definition of 
$\cM_4^{(L)}$ and $m_4^{(L)}$. } 
\beq
\label{sotto}
\cM_4 \,  =  \, 1+ \sum_{L=1}^{\infty} \cM_4^{(L)} \, = \, 
\exp \left[ \sum_{L=1}^{\infty} m_4^{(L)} \right] \ , 
\eeq
where 
\beqa
m^{(1)}_4 & = & \cM^{(1)}_4 \ ,
\\
m^{(2)}_4 & = & \cM^{(2)}_4 - {1 \over 2} \big( \cM^{(1)}_4 \big)^2 \  , 
\label{quant}
\eeqa
and so on. Motivated by  \eqref{bds} and, specifically at two loops, by \eqref{fam}, we will 
calculate in the following section the difference appearing on the right hand side of \eqref{quant}.

Let us make a final comment before moving on to explore in detail iterative structures at two loops. 
We observe that, unlike in the $\cN=4$ Yang-Mills case, the simplicity of \eqref{fullampl} does not extend immediately beyond the four-particle case,
as the explicit results for the  $n$-point amplitude of \cite{bdpr} show.
It was shown in  \cite{bdpr}, using $\cN=8$ Ward identities, that the ratio 
${\cM^{(L)} (1^+, 2^+ , \ldots , i^- , \ldots , j^- , \ldots , n^+) /
\lan i\, j\ran^8}$   is independent of the positions $i$, $j$ of the
negative-helicity gravitons,  i.e.~it is helicity blind. This is
similar to  the Yang-Mills case \cite{bddk-selfdual}, where $\cN=4$ supersymmetric Ward identities allow one to move the position of the negative-helicity particle, and show that the corresponding ratio 
in $\cN=4$ Yang-Mills
${\cM^{(L)}_{\rm YM}  (1^+, 2^+ , \ldots , i^- , \ldots , j^- , \ldots , n^+) /
\lan i\, j\ran^4}$  is independent of $i$ and $j$. 
In gravity however, this helicity-blind function  is in general expressed as  a sum of terms containing 
different  spinor bracket valued coefficients. Two immediate consequences of this we would like to stress are that, firstly, it is not immediately clear what sort of 
iterative structures could be realised beyond four points;  
and, secondly, it is not  obvious  how a Wilson loop calculation could  reproduce such terms
(this situation somewhat parallels the problems one would encounter in attempting a derivation 
of non-MHV amplitudes in $\cN=4$ super Yang-Mills from Wilson loops). 
For these reasons, in this paper we only concentrate on the four-point MHV scattering amplitudes.

\subsection{Iterative structure of the $\cN=8$ MHV amplitude at two loops}
\label{gravabdk}

The previous discussion shows that,  in searching  for prospective iterative structures 
in  the $\cN=8$ MHV amplitudes at two loops,  
it is meaningful  to analyse the quantity \eqref{quant} in
supergravity, corresponding to the two-loop term in the expansion of the logarithm 
of the amplitude. 
We will carry out
this computation in detail
for the four-point MHV gravity amplitude described in the previous subsection.
We observe that unlike the Yang-Mills
ABDK conjecture \cite{abdk}, but
in agreement with Weinberg's result for gravity
amplitudes~\cite{weinberg}, the one-loop 
infrared divergent terms of the amplitude exponentiate.  
More precisely, we will show that 
\begin{align}
  \cM_4^{(2)}- {1 \over 2} \big( \cM_4^{(1)} \big)^2 \, = \, \rm{finite}\ , 
\label{sopra}
\end{align}
and calculate the function  on the right hand side of 
\eqref{sopra}. 

The one-loop amplitude $\cM_4^{(1)}$ is given in~\eq{4pt1loop}. 
The two-loop amplitude was computed in~\cite{hep-th/9802162}, and 
is 
\beq
   {\cal M}_4^{(2)}
 \ = \
 \left({\kappa \over 2} \right)^4 \,stu\,
  \Big[s^2 \, \cI_4^{(2),{\rm P}}(s, t)
+ s^2 \, \cI_4^{(2),{\rm P}}( s,u) + s^2 \, \cI_4^{(2),{\rm NP}}(s, t)
+ s^2 \, \cI_4^{(2),{\rm NP}}(s,u)
 \ + \  \hbox{cyclic} \Big] \,.
\label{2loop}
\eeq
Here $\cI_4^{(2),{\rm P}}(s, t)$ and $\cI_4^{(2),{\rm NP}}(s, t)$ are the  planar
and non-planar double box functions:
\begin{align}
   \cI_4^{(2),  {\rm P}}(s,t) &= \int\!\!{d^{D}l\over (2\pi)^{D}} \;
 {d^{D}k\over (2\pi)^{D}} \;
 {1\over l^2 \, (l - p_1)^2 \,(l - p_1 - p_2)^2 \,(l + k)^2 k^2 \,
        (k-p_4)^2 \, (k - p_3 - p_4)^2 } \,, \cr
\cI_4^{(2) , {\rm NP}}(s,t) & = \int\!\!{d^{D} l \over (2\pi)^{D}} \,
            {d^{D} k \over (2\pi)^{D}} \
{1\over l^2\, (l-p_2)^2 \,(l+k)^2 \,(l+k+p_1)^2\,
  k^2 \, (k-p_3)^2 \, (k-p_3-p_4)^2} \, , 
\end{align}
and in \eqref{2loop} we have to sum over the three cyclic permutations of the momenta 
$p_2$, $p_3$ and $p_4$  (i.e.~over the three cyclic permutations of $s$, $t$ and  $u$). 
 
The two-loop planar box function was first evaluated
by Smirnov~\cite{hep-ph/9905323} (see also~\cite{bds}) and the non-planar double-box
function was evaluated by Tausk~\cite{hep-ph/9909506}.
These expressions need to be evaluated in different analytic regions, due to the
permutation of kinematic invariants: we fix $s,t<0$ but we will then
need functions in which $s$ or $t$ are replaced by $u=-s-t>0$,
requiring a rather delicate procedure for analytic continuation. This
procedure is outlined in Appendix~\ref{app-C}.

Smirnov's result  for the planar double box integral (we use the form given in~\cite{bds}) is given in
terms of functions
$F^{(2),{\rm P}}(s,t)$ as
\begin{align}\label{pdb}
  \cI_4^{(2) , {\rm P}}(s,t)&= \alpha_\epsilon^2 \, \left[
    {F^{(2),{\rm P}}(s,t)\over s^2 t} \right]
    \ ,\end{align}
where 
$\alpha_\epsilon:=i\,  (4 \pi)^{\epsilon-2} \Gamma(1+\epsilon)
$
and
\begin{align}
  F^{(2),{\rm P}}(s,t)= -{e^{-2\epsilon \gamma}\over \Gamma^2 (1+\epsilon)}
  (-s)^{-2\epsilon} \sum_{j=0}^4 {c_j(-t/s)\over \epsilon^j}
\ ,
\end{align}
with   the coefficients $c_j$ in (B.5) of~\cite{bds}.
This expression is valid in the region $s,t<0$ and we must carefully
analytically continue into other regions as described in Appendix~\ref{app-C}.

Tausk's expression~\cite{hep-ph/9909506} for the non-planar double box is given in terms of
functions $F^{(2),{\rm NP}}(s,t)$  as
\begin{align}\label{npdb}
 \cI_4^{(2) , {\rm NP}}(s,t)= \alpha_\epsilon^2 \, \left[
    {F^{(2),{\rm NP}}(s,t)\over s^2 t} + {F^{(2),{\rm NP}}(s,u)\over s^2 u}\right]\ .
\end{align}
The function $F^{(2), {\rm NP}}(s,t)$ is given in \cite{hep-ph/9909506}
in all analytic regions (there it is called $F_t$).

Using the above results for the integrals,
we arrive at the following expression for the two-loop amplitude,
\begin{align}
   {\cal M}_4^{(2)}
 & =  \left({\kappa^2 \alpha_\epsilon \over 4} \right)^2 \,
  \Big[su  F^{(2),{\rm P}}(s,t) + 2su F^{(2),{\rm NP}}(s,t)\nonumber\\
& \hskip 5 cm  +su F^{(2),{\rm P}}(u,t) + 2su F^{(2),{\rm NP}}(u,t) \,
 \ \ + {\mathrm{cyclic}} \Big] \,.\label{2loopf}
\end{align}
Notice that the functions $F^{(2),\mathrm{P}}(s,t)$ and
$F^{(2),\mathrm{NP}}(s,t)$ always appear together in the combination $F^{(2),\mathrm{P}}+2F^{(2),\mathrm{NP}}$, although
 $F^{(2),\mathrm{P}}(s,t)$ corresponds to the planar double box function~(\ref{pdb}),
whereas $F^{(2),\mathrm{NP}}(s,t)$ corresponds to one of the two terms in the non-planar double
box function~(\ref{npdb}).

The one-loop amplitude \eqref{4pt1loop} is expressed  as a sum of zero-mass box functions 
$ \cI_4^{(1)}$, 
where 
\begin{align}
   \cI_4^{(1)}(s,t)= \alpha_\epsilon\, \left[
     {F^{(1)}(s,t)\over s t} \right] \ ,
\end{align}
and 
\begin{align}
  F^{(1)}(s,t)= {e^{-\epsilon \gamma}\over \Gamma  (1+\epsilon)  }
  (-s)^{-\epsilon} \sum_{j=-2}^2 {\tilde{c}_j(-t/s)\over \epsilon^j}
  \ .
\end{align}
The coefficients $\tilde{c}_j$ are given in (B2) of~\cite{bds}. Again this is
valid for $s,t<0$ and we analytically continue to other regions.
Together with~(\ref{4pt1loop}), this gives the following expression for the one-loop amplitude, 
\beq
 \label{target}
 \cM^{(1)}_{4} \ = \ -i \, \left({\kappa^2 \alpha_\epsilon \over 4} \right)  \, 
 \Big[ 
 u\, F^{(1)}(s,t) +t \, F^{(1)}(s, u) +s \, F^{(1)} (u,t)
 \Big] \ .
\eeq

On putting in the functions for all permutations --  correctly defined
in their respective analytic regions --  into 
the formula for the amplitude~(\ref{2loopf}),   we
find that ${\cal M}_4^{(2)} - \half ({\cal M}_4^{(1)})^2$ is finite. 
This finite remainder is explicitly given in~(\ref{2loopfull}). 
As described in detail in Appendix \ref{app-C}, 
this function can be considerably simplified  to the following expression:%
\footnote{Notice that \eqref{finalmente} is somewhat formal, as there is no common region where all the functions appearing are away from their branch cuts. The precise analytic continuations 
for the case $s$, $t<0$ are explained in detail in Appendix C, and the explicit, somewhat lengthier  
expression for the right hand side of \eqref{finalmente} valid in that region, is given in \eqref{2loopfull}.
} 
\begin{align} 
\label{finalmente}
{\cal M}_4^{(2)} - {1\over 2}  ({\cal M}_4^{(1)})^2\ = &\  
 -\left({\kappa \over 8\pi} \right)^4 \,\Big[ 
u^2 \big[k(y)+k(1/y)\big]
  \, + \, s^2 \big[ k(1-y)+k(1/(1-y))\big]
  \nonumber\\ \cr
&\qquad \qquad \qquad +t^2 \big[k(y/(y-1))+k(1-1/y)\big] \Big]  + O(\epsilon)\ ,
\end{align}
where
\begin{align}
k(y)\ &:= 
\frac{L^4}{6}+\frac{\pi ^2 L^2}{2}-4 S_{1,2}(y) L+\frac{1}{6} \log
^4(1-y)+4 \
S_{2,2}(y)-\frac{19 \pi ^4}{90}\nonumber \\
&+i \
\left[-\frac{2}{3} \pi  \log ^3(1-y)-\frac{4}{3} \pi ^3 \log (1-y)-4 L \pi  \
\text{Li}_2(y)+4 \pi  \text{Li}_3(y)-4 \pi  \zeta (3)\right]
\end{align}
where $y=-s/t$ and $L:=\log(s/t)$.  Generalised polylogarithms, including the 
Nielsen polylogarithms $S_{m,n}$ which appear above, are discussed in 
\cite{Remiddi:1999ew}.

After  submitting this paper, we have compared our results to those of 
\cite{Naculich:2008ew}, which contains a different form 
for the finite remainder~(\ref{finalmente}). The two expressions 
are  in fact in complete agreement. Specifically,  
one can rewrite~(\ref{finalmente})
as  
\beq
{\cal M}_4^{(2)} - {1\over 2}  ({\cal M}_4^{(1)})^2\ = \left({\kappa \over 8\pi} \right)^4 \,\left[
st\, h\Big({-s\over u}\Big)+st\, h\Big(-{t\over u}\Big) + \mathrm{permutations}\right]+\cO(\epsilon) 
\ , 
\eeq
where 
\beq
h(w)\ :=\ \frac{\log ^4(w)}{3}+8 S_{1,3}(w)+\frac{4 \pi ^4}{45}+i \left[\frac{4}{3} \pi  \log ^3(w)-8 \pi  S_{1,2}(w)+8 \
\pi  \zeta (3)\right]
\ , 
\eeq
which after taking into account the 
different analytic regions considered (here we consider $s,t<0$
whereas the authors of~\cite{Naculich:2008ew} consider $s,u<0$) 
is in  precise agreement with the result of \cite{Naculich:2008ew}.

An interesting observation is that the functions appearing in the
expression for the amplitude have uniform transcendentality. 
This is somewhat surprising -- although the box function and the planar
double box function have uniform transcendentality, the non-planar
double box does not. Nevertheless,  the combination of functions 
$F^{(2),\mathrm{NP}}(s,t)+F^{(2),\mathrm{NP}}(u,t)$, which  appears 
after summing over all permutation,
does have uniform transcendentality. We notice that  amplitudes  in $\cN=1, 4$
supergravity do not have this property. 
This is explicitly shown  by the calculations in \cite{dn} of the one-loop 
four-graviton  MHV amplitudes, see Eq.~(4.6) of that paper. 
Perhaps unexpectedly,  the $\cN=6$ MHV amplitude is also maximally transcendental at one loop. 
It would be interesting to know if this property  persists at higher loops in the perturbative 
expansion of the amplitudes in these theories.

\section{The one-loop Wilson loop calculation}
In this section we describe the one-loop calculation of the four-point
MHV amplitude of gravitons from a Wilson loop.

The expression we are going to use is motivated by its application in the
eikonal approximation \cite{eik1,eik2,eik3}
to gravity \cite{ko,fpvv}, and it reads
\beq
\label{wil}
W[ \cC]  \ := \ \left\langle 
\cP \exp \left[ i \kappa \oint_{\cC} \! d\tau \  h_{\mu \nu}  (x(\tau )) \dot{x}^{\mu} (\tau )
 \dot{x}^{\nu} (\tau )   \right]\right\rangle
\ ,
\eeq
where  $ x^\mu (\tau)$ parametrises the loop $\cC$.%
\footnote{The same expression for the gravity Wilson loop has recently been used  in \cite{Green:2008kj}.}
Note that the exponent in \eqref{wil} can be rewritten as%
\footnote{In this section we set $D= 4 - 2\epsuv$.}
\beq
\int\!\!d^Dx \, \cT^{\mu \nu}(x) h_{\mu \nu}(x) \ ,
\eeq
where, in the linearised approximation, the energy-momentum tensor is
\beq
\cT^{\m \n} (x) \ := \ \int\!\!d\tau \ \dot{x}^\mu (\tau ) \dot{x}^\nu (\tau) \delta^{(D)} (x - x(\tau)) \ .
\eeq

The specific form of the contour $\cC$ we choose is  dictated by the graviton momenta
$p_1, \cdots , p_4$. In gravity there is no colour ordering -- the amplitude \eqref{4pt1loop} is a sum
over  the  permutations (1234), (1243), (1324) of the four external gravitons.
In order to match this from the Wilson loop side, 
we will therefore include the contribution of three
Wilson loops  with contours $\cC_{1234}$, $\cC_{1243}$, $\cC_{1324}$, where $\cC_{ijkl}$  is a contour made by joining the four graviton momenta $p_i$, $p_j$, $p_k$, $p_l$ in this order.
More precisely, the quantity we calculate at one loop will be
\beq
\label{WW}
W \ := \ W[ \cC_{1234}] \, W[\cC_{1243}] \,  W[ \cC_{1324}]
\ .
\eeq
Writing $W [ \cC_{ijkl}] \ :=  1 + \sum_{L=1}^{\infty} W^{(L)} [ \cC_{ijkl}] \ = \
\exp \sum_{L=1}^{\infty} w^{(L)}_{ijkl} $,
the one-loop term of \eqref{WW} is
\beq
W^{(1)} = W^{(1)}[ \cC_{1234}] \, + \,  W^{(1)}[ \cC_{1243}] \, + \, W^{(1)}[ \cC_{1324}] \,
\ .
\eeq

Before presenting the one-loop calculation, we would like to make a few preliminary comments.

{\bf 1.} One can check that
the expression in \eqref{wil} is not invariant under the gauge transformations
\beq
h_{\mu \nu} \to h_{\mu \nu} \, + \, \del_{\mu} \xi_{\nu} \, + \, \del_{\nu} \xi_{\mu} \ ,
\eeq
where $\xi^{\m} (x)$ is an arbitrary vector field. Furthermore, it is easy to see that
for contours composed of straight line segments joined
at cusps such as those
considered in this paper, the failure of gauge invariance is restricted to terms localised at the cusps.
We think it is therefore not completely surprising that the infrared divergent parts of the Wilson loop
will come out with an incorrect numerical prefactor from our calculation, compared to the finite parts, 
as we shall see below.

{\bf 2.} The expression \eqref{wil} is not
explicitly reparametrisation invariant, but it can be seen to arise from a reparametrisation
invariant expression involving an einbein $e$, by writing the action of a free,
massless particle as
\[
S \sim \int { d\tau \over e(\tau)} \,  
\dot{x}^\mu \dot{x}^\nu g_{\mu\nu}
\ .
\]
The energy momentum tensor resulting from this action is the one we use
in our definition of the Wilson line in \eqref{wil},  after gauge fixing $e=1$. 
The equation of motion for the
einbein just imposes the condition that the path of the particle is null. The
contour of the Wilson loop we use is piecewise null so that no problems can arise from reparameterisation invariance away from 
the cusps.

{\bf 3.} We note that the three contours appearing in \eqref{WW} 
are obtained by permuting the external momenta, not the
vertices. Due to the inherently non-planar character of gravity, one cannot consistently associate T-dual momenta to the external graviton momenta. For this reason, it is therefore 
unlikely that a version of dual conformal invariance might constrain the form of the amplitude here.

{\bf 4.} A different expression  for a gravity Wilson loop has been considered by Modanese
\cite{Modanese1,Modanese2},  where the right hand side of \eqref{wil} is replaced by
\beq
\label{wilmod}
\left\langle
\Tr \, \cU (\cC)
\right\rangle
\ ,
\eeq
where
\beq
\cU^{\alpha}_{\, \beta} (\cC ) \ := \
 \ \cP \exp \left[ i \kappa \oint_{\cC} \! d y^\mu \ \Gamma^{\alpha}_{\mu \beta} (y )    \right]
 \ ,
 \eeq
and  $\Gamma^{\alpha}_{\mu \beta} $ is the Christoffel connection. The quantity $\Tr \, \cU (\cC)$ has the advantage of being manifestly invariant under coordinate transformations \cite{Modanese2}. The calculation of the
 one-loop correction to   $\Tr \, \cU (\cC)$ for a closed loop has
 been considered already in  \cite{Modanese2}, and the result is  proportional to
 \beq
 \label{1lcg}
\kappa^2 \oint_{\cC} dx^{\mu_1} dy^{\mu_2} \, \lan
 \Gamma^{\alpha}_{\mu_1 \beta} (x) \Gamma^{\beta}_{\mu_2 \alpha} (y)
 \ran
 \ .
 \eeq
We refer the reader to Appendix \ref{app-D} for the details of  the  evaluation of  \eqref{1lcg}
in the linearised gravity approximation. The result is, dropping boundary terms,
\beq
\label{one-loop-chris2}
\kappa^2 \oint_{\cC} \! dx^\mu dy^\nu \, \lan \Gamma^\alpha_{\mu \beta} (x) \Gamma^\beta_{\nu \alpha} (y) \ran
 \ = \   c(D) \oint_{\cC}  \! dx_\mu dy^\mu \, \delta^{(D)} (x-y)
 \ ,
  \eeq
where $c(D)$ is a numerical constant which is  finite as $D\to 4$.
Parameterising the contour as $x= x(\s)$, we can rewrite the right hand side of \eqref{one-loop-chris2}
as
\beq
\label{divv}
c(D) \int\! d\tau \int\! d\sigma  \ \dot{x}_\mu (\t)  \dot{x}^\mu (\s ) \ \delta^{(D)} (x(\t ) - x(\s ))
\ .
\eeq
Some  readers may notice   that  the  divergent expression in \eqref{divv} 
already appears in the lowest-order expansion of the
Makeenko-Migdal loop equation \cite{mm1,mm2} in Yang-Mills. 
An evaluation of  \eqref{divv}  has been carried out
in $\cN=4$ super Yang-Mills in \cite{dgo} for a cusped contour, by using a regularisation of the  Dirac delta function which employs
a  cutoff of width $a$. Interestingly, the right hand side of \eqref{divv} is then found to be proportional
to  $1/a^2$ times the one-loop cusp anomalous dimension.%
\footnote{In \cite{mos} this result was extended to two loops, and conjectured to hold to any loop
order in perturbation theory.} 
We observe that, because of  the delta function appearing in  it, the expression in \eqref{divv}
receives contribution only from cusps and self-intersections present in the contour.
The main point we would like to make  here is  that \eqref{divv} does not reproduce
(parts or all  of) the $\cN=8$ supergravity amplitude  \eqref{4pt1loop}.  Therefore,  in the following
we will work with the Wilson loop defined for a polygonal contour as in \eqref{wil}.

We now proceed  to describe the calculation. We work in the de Donder gauge, where the propagator is 
given by 
\beq
\label{don}
\lan h_{\mu_1 \mu_2} (x) h_{\nu_1\nu_2} (0)\ran \ = \
{1\over 2} \Big( \eta_{\mu_1 \nu_1} \eta_{\mu_2 \nu_2} + \eta_{\mu_1 \nu_2} \eta_{\mu_2 \nu_1} -
{2\over D-2} \eta_{\mu_1 \mu_2} \eta_{\nu_1 \nu_2} \Big) \Delta (x)
\ ,
\eeq
where
\beqa
\D  (x) & := &
- {\pi^{2 - {D\over 2}} \over 4 \pi^2}
\Gamma \Big({D\over 2} - 1 \Big)
 {1\over (-x^2+ i \varepsilon)^{{D\over 2} - 1}}
\\ \nonumber
&=&
- {\pi^{\epsuv} \over 4 \pi^2} {
\Gamma (1-\epsuv)  \over
(-x^2+ i \varepsilon)^{1-\epsuv }}
\ .
\eeqa
The gravity calculation  is very similar to the one-loop calculation performed in
\cite{dks, bht} for the  one-loop Wilson loop in maximally supersymmetric Yang-Mills theory.
As in that case, three different classes of diagrams contribute at  one loop.%
\footnote{For a Wilson loop bounded by gravitons, only gravitons can be  exchanged
to  one-loop order.}
In the first one,  a graviton stretches between points belonging to the same segment.
As in the Yang-Mills calculation, these diagrams give a vanishing contribution since the momenta of the gravitons are null.
In the second class of diagrams, a graviton stretches between two adjacent segments meeting at a cusp.
In the Yang-Mills case, such diagrams  lead to ultraviolet divergences
\cite{polyakov,cu1,cu2,cu3,cu4,cu5,cu6,cu7}. 
As in the Yang-Mills Wilson loop case  \cite{dks},
these divergences are  associated with infrared divergences 
of the amplitude by identifying $\epsuv = -\eps$.

We will now see how in our gravity calculation, these divergences
are still present but will be softened (from $1/ \epsuv^{2}$ to $1/ \epsuv$)  
after taking into account the sum over
the contributions of the three Wilson loops.

\begin{figure}[ht]
\begin{center}
\scalebox{0.40}{\includegraphics{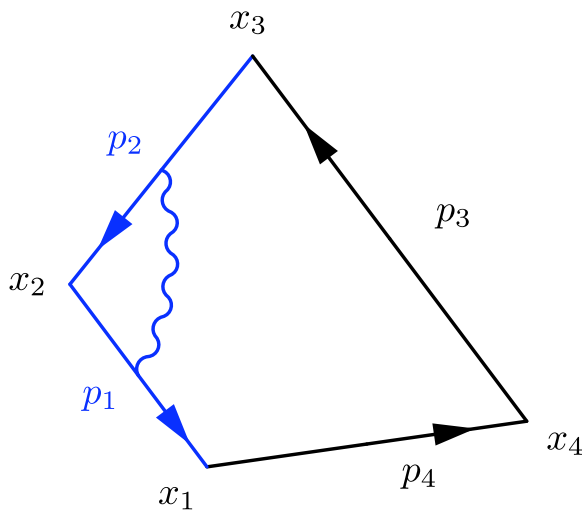}}
\end{center}
\caption{\it
A one-loop correction to the Wilson loop bounded by momenta $p_1,\cdots , p_4$,  where  a graviton
is exchanged between two lightlike momenta meeting at a cusp.
Diagrams in this class generate infrared-divergent contributions to the four-point amplitude which,
after summing over the appropriate permutations give rise to  \eqref{MIR}.
}
\label{figure1}
\end{figure}

A typical diagram in the second class is pictured in Figure \ref{figure1}.
There one has $x_1 (\tau_1) - x_2 (\tau_2 ) = p_1 (1 - \tau_1 ) + p_2 \tau_2$.
The cusp diagram gives
\beqa
\label{cuspintegral}
&&-(i \kappa \tilde \mu^{\epsuv})^2 \, {\Gamma (1 - \epsuv) \over 4 \pi^{2 - \epsuv}} \
\int_{0}^{1}\!\!d\tau_1 d\tau_2 \ {(p_1 p_2 )^2 \over [ - \big(p_1 \tau_1 + p_2 \tau_2\big)^2]^{1-\epsuv}}
\nonumber \\
&  = &
 -(i \kappa \tilde \mu^{\epsuv})^2 \, {\Gamma (1 - \epsuv) \over 4 \pi^{2 - \epsuv}} \
 \left[ {1\over 4} {(-s)^{1+ \epsuv}\over \epsuv^2 } \right] \ .
\eeqa
Notice that we need to choose $\epsuv >0$ in order to regulate the divergence
in \eqref{cuspintegral}. Furthermore the scale used in the Wilson loop
calculation is related to the scale used to regulate the amplitudes $\mu$ as
$ \tilde \mu=(c \mu)^{-1}$ (the precise coefficient $c$ in front of $\mu$  can be reabsorbed into an
appropriate redefinition of the coupling constant).

Summing this over the four cusps of the first Wilson loop, one gets%
\footnote{We set $c (\epsuv)  =  ( \kappa \tilde \mu^{\epsuv})^2 \,
{\Gamma (1 - \epsuv) /  (4 \pi^{2 - \epsuv})}$.}
\beq
{c(\epsuv) \over 2 \epsuv^2} \Big[ (-s)^{1 + \epsuv} + (-t)^{1 + \epsuv} \Big]
\ .
\eeq
Adding  the contributions of the two other Wilson loops, we get
\beq
\label{MIR}
{c(\epsuv) \over  \epsuv^2} \Big[ (-s)^{1 + \epsuv} + (-t)^{1 + \epsuv} +  (-u)^{1 + \epsuv} \Big]
\ .
\eeq
Upon expanding this expression in $\epsuv$,  the cancellation of the $1/ \epsuv^2$ pole becomes manifest
(after using $s + t + u = 0$), and   \eqref{MIR} becomes, up to terms vanishing as $\epsuv\to 0$,
\beq
- c(\epsuv)  \Big[ {1 \over \epsuv} \Big( s \log(-s)+t \log(-t)+u \log(-u)\Big)
+ {1 \over 2} \Big(s \log^2(-s)\, +t \log^2(-t)\, +u \log^2(-u)\Big) \Big] \ .
\eeq
We recognise that  this expression is the infrared-divergent part  of the four-point MHV gravity amplitude
\eqref{4pt1loop}. We notice however that, after summing over the appropriate permutations 
as in \eqref{WW}, one finds that these infrared-divergent terms have an extra factor of 
$-2$ compared to the finite parts, to be calculated below. 
We believe this mismatch is not unexpected,  given that the failure of gauge invariance 
of \eqref{wil} occurs at the cusps.%
\footnote{A factor of $2$  could be explained because we are effectively double-counting the cusps 
in summing over the permutations, however at the moment we are unable to explain the relative minus sign.}

\begin{figure}[ht]
\begin{center}
\scalebox{0.40}{\includegraphics{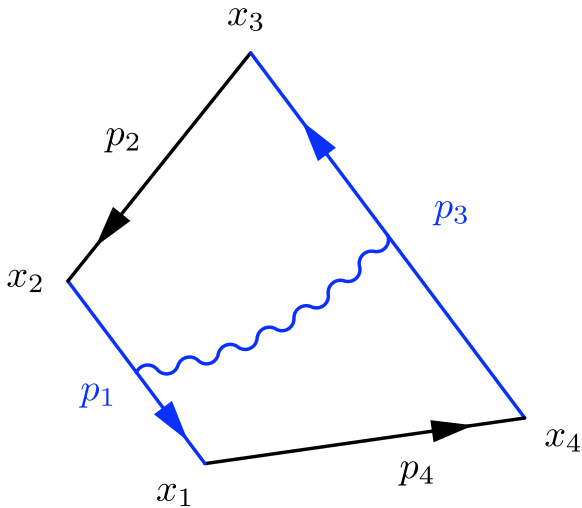}}
\end{center}
\caption{\it
Diagrams in this class, where a graviton stretches between  two non-adjacent edges of the loop, 
are finite, and give in the four-point case a contribution equal to the finite part of the zero-mass box function
$F^{(1)} (s,t)$ multiplied by  $u$. 
}
\label{figure2}
\end{figure}

We now move on to the last class of diagrams, where a graviton is exchanged between two
non-adjacent edges with momenta $p$ and $q$; one such  example is depicted in 
Figure \ref{figure2}. In the Yang-Mills case these diagrams  were found to
be in one-to-one correspondence with the finite part of the two-mass easy 
box functions with massless legs $p$ and $q$. We will show now that \eqref{wil} 
leads exactly to the same kind of correspondence
with the finite part of the one-loop four-graviton amplitude.

Indeed, the one-loop diagram in Figure \ref{figure2} is equal to
\beqa
\label{finiteintegral}
&&  c(\epsuv)
\int_{0}^{1}\!\!d\tau_1 d\tau_2 \ {(p_1 p_3 )^2 \over [ - \big(p_1 (1- \tau_1) +
p_2 + p_3 \tau_2\big)^2]^{1-\epsuv}}
\ . 
\eeqa
This integral is finite in four dimensions, and gives
\beq
c(\epsuv) \, {u \over 2} \,  {1 \over 4} \Big[ \log^2\Big({s \over t}\Big) + \pi^2 \Big] \ .
\eeq
Summing over the two possible pairs of non-adjacent segments and including the contributions of the two
other Wilson loop configurations, we get exactly the finite part of the one-loop MHV amplitude 
in  $\mathcal{N}=8$ supergravity \eqref{abm} up to the tree-level amplitude.%
\footnote{A Wilson loop calculation clearly cannot produce any dependence on  helicities and/or 
spinor brackets.  Incidentally, we also observe that in Yang-Mills, a Wilson loop calculation cannot
produce any parity-odd terms such as those appearing in the five- and six-point 
two-loop MHV amplitudes.}

\section{Calculation in the conformal  gauge}

The gravity Wilson loop defined above, unlike the Yang-Mills Wilson
loop, is gauge dependent. It turns out that one can define a gauge in
both cases in which the cusp diagrams vanish completely. We call these ``conformal" gauges.%
\footnote{This name is motivated by the fact that, in the Yang-Mills case, 
the $D$-dimensional propagator  turns out to be proportional to the inversion tensor 
$J_{\m \n}(x) := \eta_{\mu \nu} - 2 x_\m x_\n / x^2$. The Yang-Mills conformal propagator is described in 
Appendix \ref{app-A}, where we show that it can be obtained from a Feynman-'t Hooft gauge-fixing term 
with a specific coefficient.  In Appendix \ref{app-B} we perform the calculation of the 
$n$-point polygonal Wilson loop. The outcome of this calculation is that cusp diagrams
in the conformal gauge vanish, and the $\cN=4$ amplitude is obtained from 
summing over diagrams where a gluon connects non-adjacent edges. In this case, 
each such diagram is in one-to-one correspondence with a {\it complete}  two-mass easy box function.}  
In the Yang-Mills Wilson loop one obtains the same answer in either gauge,
but in the gravity Wilson loop the conformal gauge appears to be the
unique gauge which gives the amplitude, both infrared-divergent and
finite pieces correctly, to all orders in $\epsilon$.

\subsection{Gravity propagator in general gauges}
\label{gengauge}

We first need to define a general class of gauges in the gravity
case. To do this, we consider
the free Lagrangian  of linearised gravity:
\begin{align}
  {\cal L}  =&  -\frac{1}{2} (\de_\mu h_{\nu\rho})^2 + (\de_\nu
  h^\nu{}_\mu )^2 +\frac{1}{2} (\de_\mu h^\lambda{}_\lambda)^2 + h^\lambda{}_\lambda
  \de_\mu \de_\nu h^{\mu\nu}
  \ ,
\end{align}
which can be easily checked to be invariant with respect to
the gauge transformation $ \delta h_{\mu\nu} = 2 \de_{(\mu} \xi_{\nu)}$.
We then add a gauge fixing term of the following form:
\begin{align}
\label{ddlgft}
  {\cal L}^{(\rm{gf})} =&  \frac{\alpha}{2} \Big(\de_\nu h^{\nu}_\mu-{1 \over
    2}\de_\mu h^\alpha_\alpha\Big)^2\ ,
\end{align}
which is de Donder-like, but with an arbitrary  free parameter $\alpha$. We will
call this the $\alpha$-gauge.

In momentum space, the corresponding  gauge-fixed Lagrangian has the form
$  (1/2) h^{\mu\nu} K_{\mu\nu,\mu'\nu'} h^{\mu'\nu'}
$,
where
\beqa
  K_{\mu\nu,\mu'\nu'}(k)& = & k^2
  \eta_{\mu'(\mu}\eta_{\nu)\nu'}-2k_{(\mu}\eta_{\nu)(\nu'}
    k_{\mu')}- k^2\eta_{\mu\nu}\eta_{\mu'\nu'} +\eta_{\mu\nu}k_{\mu'}
    k_{\nu'}+ \eta_{\mu'\nu'}k_{\mu}k_{\nu}\nonumber \\
&-& \alpha\Big[  k_{(\mu} \eta_{\nu)(\nu'}
    k_{\mu')} -{1\over2}(\eta_{\mu\nu}k_{\mu'}
    k_{\nu'}+ \eta_{\mu'\nu'}k_{\mu}k_{\nu})+{1\over 4} k^2\eta_{\mu\nu}\eta_{\mu'\nu'}\Big]
\ .
\eeqa
Now we define the propagator $D_{\mu\nu,\nu'\mu'}$ to be the inverse of
$K_{\mu\nu,\mu'\nu'}$, i.e.
\begin{align}
  K_{\mu\nu,\mu'\nu'}D^{\mu'\nu',mn}=\delta^{(m}_{\mu}\delta^{n)}_{\nu}\ .\label{KD=1}
\end{align}
By writing down the most general Lorentz covariant terms which have
the correct index symmetries and have mass dimension equal to -2,
we see that $D_{\mu\nu,\mu'\nu'}$ must take the form
\beqa
  D_{\mu\nu,\mu'\nu'}(k)&=&{1 \over k^2} \eta_{\mu'(\mu}\eta_{\nu)\nu'}
  + {a\over k^4}  k_{(\mu} \eta_{\nu)(\nu'}
    k_{\mu')} +{b\over k^2}\eta_{\mu\nu}\eta_{\mu'\nu'} \nonumber \\&+& {c\over k^4}
    (\eta_{\mu\nu}k_{\mu'}
    k_{\nu'}+ \eta_{\mu'\nu'}k_{\mu}k_{\nu}) + {d\over k^6} k_\mu
    k_\nu k_{\mu'} k_{\nu'}
    \ .
\eeqa
Then \eqref{KD=1} gives a set of equations for the free
parameters which have the unique solution (for $D\neq2$),\quad 
$a\!=\!-(4+2\alpha)/\alpha$,\quad  
$b\!=\!-1/(D-2)$,\quad   $c\!=\!d\!=\!0$. Thus, the propagator corresponding to the
$\a$-gauge defined by the gauge-fixing term
\eqref{ddlgft}  is given by
 \begin{align}
 \label{cpg}
  D_{\mu\nu,\mu'\nu'}(k)={1 \over k^2} \Big(\eta_{\mu'(\mu}\eta_{\nu)\nu'}
-{1\over D-2}\eta_{\mu\nu}\eta_{\mu'\nu'}\Big)
  -{4+2\alpha\over \alpha}  {1\over k^4}  k_{(\mu} \eta_{\nu)(\nu'}
    k_{\mu')}
    \ .
\end{align}
Notice that \eqref{cpg}  reproduces the standard de Donder propagator for $\alpha=-2$.

\subsection{Propagator in position space in the $\alpha$-gauge}

We now perform the Fourier transform to position space.
The Fourier transform of $1/k^{2\lambda}$ has
the form%
\footnote{More details can be found in Appendix \ref{app-A}.}
\begin{align}\label{fl}
 \cF[1/k^{2\lambda}]=c(D,\lambda)(-x^2)^{\lambda-D/2}
 \ ,
\end{align}
where 
\begin{align}
 c(D,\lambda)=-4^{-\lambda}\pi^{-D/2} {\Gamma(2-D/2)\Gamma(D/2-1)\over \Gamma(\lambda+1-D/2)\Gamma(\lambda)}
 \ . 
\end{align} 
By differentiating  twice with respect to  $x$ and setting $\lambda=2$
we find that
the Fourier transform of $k_\mu k_\nu/k^4$ is
\begin{align}\label{ftkk}
2c(D,2) \epsuv \left[{\eta_{\mu\nu}  \over (-x^2)^{1-\epsuv}}+{2x_\mu
x_\nu \over (-x^2)^{2-\epsuv}} (1-\epsuv)\right]\ .
\end{align}
Using this we take  the Fourier transform of 
\eqref{cpg}, and obtain 
the propagator in position space:
\beq
D_{\mu\nu,\mu'\nu'}(x)
\, =\,
A  {\eta_{\mu'(\mu}\eta_{\nu)\nu'} \over (-x^2)^{1-\epsuv}} -{c(D,1)\over D-2}{1 \over (-x^2)^{1-\epsuv}}\eta_{\mu\nu}\eta_{\mu'\nu'}
  + B {1\over (-x^2)^{2-\epsuv}}  x_{(\mu} \eta_{\nu)(\nu'}
    x_{\mu')}\ ,
\eeq
where
\begin{align}
  A= c(D,1)+2 a \,\epsuv \,c(D,2) \qquad \qquad B=4a \,\epsuv
  (1-\epsuv)\, c(D,2)\ ,
\end{align}
and
\begin{align}
  a=-{4+2\alpha\over \alpha}
  \  .
\end{align}

\subsection{The conformal gauge}

By direct analogy with the Yang-Mills case, discussed in Appendix \ref{app-B}, 
where we show that in the ``conformal" gauge the cusp diagrams vanish, 
we define  the gravity  conformal gauge to be the gauge
in which the cusp diagrams vanish. We show in this section that this particular gauge can be obtained
from an $\a$-gauge fixing term as defined in the previous section for an appropriate
value of the parameter $\a$.

To begin with, consider the cusp defined by momenta
$p,q$ and then let $x=p \sigma+q \tau$.  Then the term appearing
in the cusp at one loop is
\begin{align}
  p^\mu p^\nu D_{\mu\nu,\mu'\nu'}(x) q^{\mu'} q^{\nu'}=(-x^2)^{\epsuv-2}(pq)^3\sigma\tau(B-2A)
  \ .
\end{align}
Therefore, the cusp diagrams vanish for  $B=2A$. One can quickly check that this
implies
 $a=-c(D,1)/(2\epsuv^2c(D,2))=4/(D-4)$.
The result is the
 propagator in the conformal gauge:
 \beqa
 D_{\mu\nu,\mu'\nu'}(x)&=&
 c(D,1){\epsuv-1
  \over \epsuv} \bigg[   {1 \over (-x^2)^{1-\epsuv}}
\Big(\eta_{\mu'(\mu}\eta_{\nu)\nu'}+{\epsuv\over
  2(\epsuv-1)^2}\eta_{\mu\nu}\eta_{\mu'\nu'} \Big)\nonumber \\
& +&
  2 {1\over (-x^2)^{2-\epsuv}}  x_{(\mu} \eta_{\nu)(\nu'}
    x_{\mu')}\bigg] \ ,
 \eeqa
which requires
\begin{align}
  \alpha=-2(D-4)/(D-2)\ .
\end{align}

\subsection{Gravity Wilson loop in the conformal gauge}

We now proceed to calculate the gravity Wilson loop in this conformal
gauge. We have shown  that the cusp diagrams are equal to zero in this gauge,  
therefore we need only calculate the ``finite'' diagrams 
(which are now no longer finite). Consider the Wilson loop with
edges $p_1,p_2,p_3,p_4$ (in that order) and the graviton stretching
between sides 1 and 3. Then we have $x=\sigma p_1+ \tau p_3 +p_2$ and
$x^2= s \sigma+t \tau +u \sigma \tau$. The contribution of this diagram is then
\begin{align}
&  \int_0^1 \!d\sigma d\tau\ {p_1}^\mu {p_1}^\nu D_{\mu\nu,\mu'\nu'}(x) {p_3}^{\mu'}
  {p_3}^{\nu'} 
  \\
=\, &\ c(D,1)\, {\epsuv-1 \over \epsuv} {u\over 4} \int_0^1 d\sigma d\tau
{s t \over (-(s \sigma+t \tau +u \sigma \tau))^{2-\epsuv}}\nonumber \\
=\, 
\nonumber
&\ c(D,1)\, {1 \over \epsuv^2} {u\over 4} \Big[-(-s)^\epsuv
{}_2F_1(1,\epsuv,1+\epsuv,1+{s\over t} )-
(-t)^\epsuv{}_2F_1(1,\epsuv,1+\epsuv,1+{t\over s})\Big]
\ . 
\nonumber  
\end{align}
We see that we obtain the complete (infrared-divergent as well as finite
pieces) two-mass easy box function to all
orders in $\epsuv$.
Adding the other diagram (which gives the same result)  and then
summing over the remaining permutations as described
above, gives the correct one loop
$\cN=8$ supergravity amplitude~\eq{target}.

Despite this encouraging result, we should remember 
that our starting expression for the Wilson loop \eqref{wil} 
was not gauge invariant. It would be important to remedy  
this gauge non-invariance, which is localised at the positions of 
the cusps, by an appropriate subtraction procedure. 
Furthermore, it would be interesting to study infrared divergences, as well as the 
derivation of finite parts 
of gravity amplitudes at higher loops using the Wilson loop
proposed in \eqref{WW}.

\vspace{1.4cm}


\section*{Acknowledgements}

It is a pleasure to thank  James Bedford, Lance Dixon, Valeria Gili,
Marco Matone, Sanjaye Ramgoolam, Costas Zoubos and especially
George Georgiou  for discussions. We would also like to thank Lance Dixon 
for letting us know that his result for \eqref{finalmente}  agrees with ours.  
This work was supported by the STFC under a Rolling Grant PP/D507323/1.
The work of PH is supported by an EPSRC Standard Research Grant EP/C544250/1.
GT is supported by an EPSRC Advanced Research Fellowship EP/C544242/1
and by an EPSRC Standard Research Grant EP/C544250/1.

\vspace{1cm}

\newpage
\appendix

\section{The conformal propagator in  Yang-Mills} \label{app-A}
In this section we briefly outline the construction of the conformal propagator.
It is defined to be proportional to the inversion tensor
\beq
J_{\mu \nu} (x) := \eta_{\mu \nu} \, - \, 2 {x_\m x_\n \over x^2}
\ .
\eeq
By using
\beqa
\int\! {d^Dp \over (2 \pi)^D}e^{i p x} \ { 1\over p^2}  & = &-  {\pi^{-{D\over 2} }\over 4 } \Gamma \Big( {D\over 2} - 1 \Big)
{1 \over ( - x^2 + i \varepsilon)^{ {D\over2} - 1} }
 \ ,
 \\ \nonumber
\int\! {d^Dp \over (2 \pi)^D}e^{i p x} \ { p_\m p_\n\over p^4}  & = &- {\pi^{-{D\over 2} }\over 8 } \Gamma \Big( {D\over 2} - 1 \Big)
{\eta_{\m \n} - (D-2) x_\m x_\n / x^2 \over ( - x^2 + i \varepsilon)^{ {D\over2} - 1} }
\ ,
  \eeqa
  it is easy to see that the following combination has the desired property:
  \beq
  \int\! {d^Dp \over (2 \pi)^D}e^{i p x}\ {\eta_{\m \n} \over p^2} \ + \
{4\over D-4} \, \int\! {d^Dp \over (2 \pi)^D}e^{i p x}\ {p_\m p_\n  \over p^4}  \ = \
\Delta^{\rm conf}_{\m \n}  (x) \ ,
\eeq
where we define the conformal propagator
\beq
\label{confpropa}
\Delta^{\rm conf}_{\m \n} (x) \ := \ -{D-2 \over D-4}  {\pi^{-{D\over 2} }\over 4 } {\Gamma \Big( {D\over 2} - 1 \Big) \over
( - x^2 + i \varepsilon)^{ {D\over2} - 1} } \, \Big[ \eta_{\m \n} \, -\,  2 {x_\m x_\n \over x^2} \Big]
\ .
\eeq
Thus, the expression  \eqref{confpropa} is obtained by choosing a Feynman-'t Hooft gauge-fixing term
 $(\alpha / 2) \int\! d^Dx \,  (\del_\mu A^\mu)^2$ for the particular  choice of  $\alpha = (D-4)/D$. 
The vanishing of this gauge-fixing term in $D=4$ dimensions  is reflected in the presence of a 
factor of $1/(D-4)$ in \eqref{confpropa}, 
which makes this propagator not well defined  in four dimensions.

\section{The Yang-Mills Wilson loop with  the conformal propagator}\label{app-B}
\begin{figure}[ht]
\begin{center}
\scalebox{0.39}{\includegraphics{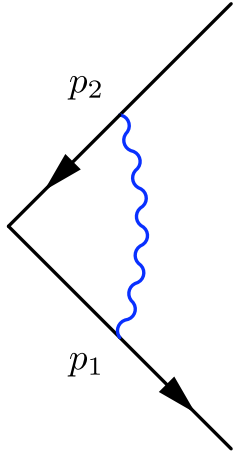}}
\end{center}
\caption{\it
A one-loop correction for a cusped  contour.
We show in the text that, when evaluated in the conformal gauge, the result of this diagram
vanishes.
}
\label{cuspfig}
\end{figure}

As a simple but illuminating  application of the above conformal propagator, we  would like
to outline the calculation of the Yang-Mills Wilson loop with a  contour made of
$n$ lightlike segments performed in \cite{bht}.
Of course, the usual expression of the Wilson loop in Yang-Mills
is gauge invariant, hence evaluating it in any gauge leads to the same result.
The use of this gauge leads however to a recombination of terms,
where the cusp diagrams vanish.%
\footnote{The usual infrared-divergent terms are produced by diagrams  which, in the Feynman gauge
calculation of \cite{bht}, were finite.}
Consider for instance the cusped contour  depicted  in Figure \ref{cuspfig}.
Using the conformal propagator, and $x_{p_1} (\tau_1) - x_{p_2} (\tau_2) = p_1 (1 - \tau_1) + p_2 \tau_2$,
we see that the one-loop correction to the cusp
is given by an expression proportional to
\beq
\int\!d\tau_1 d\tau_2\
p_{1 \mu} p_{2\nu}
{
\eta^{\mu \nu} -
2 {[ p_1 (1-\tau_1) + p_2 \tau_2]^\m
    [ p_1 (1-\tau_1) + p_2 \tau_2]^\n
    \over 2 (p_1 p_2) (1-\tau_1) \tau_2}
    \over
[ - 2 (p_1 p_2) (1-\tau_1) \tau_2 ]^{D/2 - 1} }
\ ,
\eeq
which vanishes.

\begin{figure}[ht]
\begin{center}
\scalebox{0.33}{\includegraphics{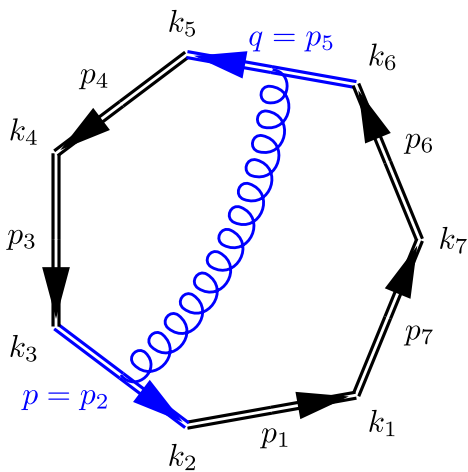}}
\end{center}
\caption{\it
A one-loop diagram where a gluon connects two non-adjacent segments.
In the Feynman gauge employed in \cite{bht}, the result of this diagram is equal to the finite part of
a two-mass easy box function $F^{\rm 2me} (p, q, P, Q)$,
where $p$ and $q$ are the massless legs of the two-mass easy box, and correspond to the segments
which are connected by the gluon. In the conformal gauge, this diagram is equal to the full
box function. The diagram depends on the other gluon momenta only through the
combinations $P$ and $Q$. In this example, $P=p_3 + p_4$, $Q= p_6 + p_7 + p_1$.  }
\label{n-finite}
\end{figure}

We now move on to consider diagrams where a gluon is exchanged between non-adjacent segments,
such as that in Figure \ref{n-finite}.
In \cite{bht} it was shown that this diagram is equal to the finite part of a two-mass easy box function.
In the conformal gauge, a simple calculation shows that it is equal to%
\footnote{In the following we set $\eps = - \epsuv$.}
\beq
\label{se-che}
f_\eps \cdot {1\over2} (s t \, - \, P^2 Q^2) \, \int_{0}^{1} {d\tau_1 \, d\tau_2 \over [
-D(\tau_1 , \tau_2)]^{2 + \eps}} \ ,
\eeq
where
\beqa
D (\tau_1 , \tau_2) & := & (x_p (\tau_1) - (x_q (\tau_2))^2
\\ \nonumber
&=& P^2 + (s-P^2) (1 - \tau_1) -  (t - P^2) \tau_2 - u (1 - \t_1) \t_2
 \ ,
 \eeqa
where we used $2 (pP) = s - P^2$, $2 (q P) = t - P^2$, and
$ s + t + u = P^2 + Q^2$. We have also introduced
\beq
f_\eps \ := \ {1 + \eps \over \eps} {\Gamma (1 + \eps)  \over  \pi^{2 + \eps}}
\ .
\eeq
In  \cite{bht} it was found that
\beq
\label{moz}
\int_{0}^{1} {d\tau_1 \, d\tau_2 \over [
-D(\tau_1 , \tau_2)]^{2 + \eps}} \ = \ {\cF_{\eps + 1} \over P^2 + Q^2 - s - t}
\ ,
\eeq
where
\beqa
\label{333}
&&\mathcal{F}_{\epsilon} \, = \,
-{ 1 \over \e^2}
\\ [6pt]\nonumber
&&
\hspace{-0.4cm}
\cdot
\left[
\Big( {a \over 1-aP^2} \Big)^{\e} \,
\mbox{}_{2}F_1 \left( \e, \e, 1+ \e, {1 \over 1 - a P^2 }\right)
\, + \,
\Big( {a  \over 1-aQ^2} \Big)^{\e} \,
\mbox{}_{2}F_1 \left( \e, \e, 1+ \e, {1 \over 1 - a Q^2 }\right)
\right.
\\ [6pt] \cr
&& \hspace{-0.4cm}  - \,
\left.
\Big( {a  \over 1-as} \Big)^{\e} \,
\mbox{}_{2}F_1 \left( \e, \e, 1+ \e, {1 \over 1 - a s }\right)
\, - \,
 \Big( {a  \over 1-at} \Big)^{\e} \,
\mbox{}_{2}F_1 \left( \e, \e, 1+ \e, {1 \over 1 - a t }\right)
\right]
\ , 
\nonumber
\eeqa
where we have introduced 
\beq
a := {P^2 + Q^2 - s - t \over P^2 Q^2 - s t }
 \ .
\eeq
Notice that in \eqref{moz} the function $\cF$ appears with argument $\eps + 1$.
After a moderate use of hypergeometric identities, we find that the one-loop correction in
\eqref{se-che} is equal to
\beq
 {1 \over 2} {\Gamma ( 1 + \eps ) \over 4 \pi^{2+ \eps}  } \ F^{\rm 2me} (s, t, P^2, Q^2) \ ,
\eeq
where $F^{\rm 2me} (s, t, P^2, Q^2)$ is the all-orders in $\eps$ expression of the two-mass easy box function
derived in \cite{ftt},%
\footnote{Omitting a factor of  $ c_\G \ = \ \G (1 + \e) \G^2 ( 1 -  \e) /  (4\pi)^{2- \eps} $ compared to \cite{ftt}.}
\beqa
\nonumber
&& \hspace{-0.7cm}
F^{\rm 2me} (s, t, P^2, Q^2) =
-{ 1 \over \e^2}
\left[
 \Big( {-s \over \mu^2} \Big)^{-\e} \,
\mbox{}_{2}F_1 \left( 1, -\e, 1- \e, as \right)
\, + \,
\Big( {-t \over \mu^2} \Big)^{-\e}
\mbox{}_{2}F_1 \left( 1, -\e, 1- \e, at \right)
\right.
\\  [6pt]\cr
&&  \qquad - \,
\left.\Big( {-P^2 \over \mu^2} \Big)^{-\e}
\,
\mbox{}_{2}F_1 \left( 1, -\e, 1- \e, aP^2 \right)
\, - \,
 \Big( {-Q^2 \over \mu^2} \Big)^{-\e} \,
\mbox{}_{2}F_1 \left( 1, -\e, 1- \e, a Q^2 \right)
\right]
\label{puzzola}
\eeqa
Summing over all possible gluon contractions in the Wilson loop, one finds complete agreement with
the result derived in \cite{bht} for the same Wilson loop, as anticipated.

\section{Analytic continuation of two-loop box functions}\label{app-C}

In Section~\ref{sec:2loops} the one and two loop amplitudes are given 
in terms of functions $F^{(2),\mathrm{P}}(s,t)$, $F^{(2),\mathrm{NP}}(s,t)$ and
$F^{(1)}(s,t)$. In Yang-Mills,  colour ordering means that we need to define the functions
explicitly in only one
analytic regime.
In gravity however,  we must sum over permutations of the
kinematic invariants.
Even if we fix the kinematic regime to be $s,t<0$ we must also
consider for example  $F(s,u)$,  and the second argument 
of this function will be greater than zero (recall that $u=-s-t$).
There will be three different kinematic regimes of interest and, 
following Tausk~\cite{hep-ph/9909506},  we label them in the following
way:
\begin{align}
  F(s,t)=\left\{
    \begin{array}[c]{lllcl}
      F_1(s,t) \qquad t,u<0 \qquad &\\
      F_2(s,t) \qquad s,u<0 \qquad &\\
      F_3(s,t) \qquad s,t<0 \ \ . \qquad &
    \end{array}
  \right.
\end{align}

Tausk gives explicit formulae for the non-planar box function in all
three regions, but it is nevertheless useful to know how to  obtain the
function in any region from its manifestation in a particular 
region. The Mathematica package HPL~\cite{Maitre:2007kp} is very useful for this.

We will sketch the procedure below.
Let us begin by considering the analytic continuation from region 1 to
2. In general, functions in this region take the following form:
\begin{align}\label{ff}
  F_1(s,t)=f(\log(s),\log(-t),\log(-u),H_{\vec{a},1}(-t/s))\ .
\end{align}
Here $H_{\vec{a},1}(z)$ is a harmonic polylogarithm where
$\vec{a}$ represents a string of zeros or ones. Note that at two loops
we need not use harmonic polylogarithms as they can all be 
re-expressed in terms of 
Nielsen polylogarithms. On the other hand, at higher loops harmonic 
polylogarithms will appear which cannot be so expressed; 
it is nevertheless useful to use harmonic  polylogarithms even here   
(see~\cite{Remiddi:1999ew, Maitre:2007kp} for
more details on harmonic polylogarithms). 
Such a  harmonic
polylogarithm is analytic everywhere on the complex plane except for a
branch cut on the real axis for $z>1$. Note that the
arguments of all the (poly)logarithm functions in~\eq{ff} lie away from the
branch cut.

Now the function continued to region 2 takes the following form:
\begin{align}\label{ff2}
  F_2(s,t)=f(\log(-s)+i\pi,\log(t)-i\pi,\log(-u),H_{\vec{a},1}(-t/s))\ .
\end{align}
We have analytically continued the $\log$s appropriately, however the
argument of the HPL functions now lies on the branch cut in region
2 ($-t/s=1+u/s>1$). We use the HPL package to transform away from
the branch cut. Specifically putting $-t/s=1/y$ the command
`HPLConvertToSimplerArgument' will rewrite this in terms of HPLs with
the argument $y=-s/t$ which lies off the branch cut (one must also use
the command
`HPLReduceToMinimalSet' to write the functions in a standard form).

If we wish to obtain the formula in  region 3 from that in region 1 we
immediately have a problem. The argument of our HPL functions is
$-t/s$ which is not on a branch cut for  either region. However, 
close examination shows that
as we pass smoothly from region 1 to region 3, we must  first pass along the
branch cut -- for example  we must pass through the point $s=0$,
i.e.~$-t/s=\infty$. The HPL programme will not take this into account
and the naive analytic continuation gives the wrong result. 
So it is better to first perform a transformation $y\rightarrow
1-y$ on the HPLs in $F_1(s,t)$ to find a new expression for $F_1(s,t)$ in
terms of HPLs with argument $1+t/s=-u/s$, i.e.
\begin{align}
   F_1(s,t)=g(\log(s),\log(-t),\log(-u),H_{\vec{a},1}(-u/s))\ .
\end{align}
Then in region 3 we find $-u/s>0$,  and hence we are on the branch cut and we can
proceed as before. We analytically continue as follows,
\begin{align}
  F_3(s,t)=g(\log(-s)+i\pi,\log(-t),\log(-u)-i\pi ,H_{\vec{a},1}(-u/s))\ .
\end{align}
Now use the HPL programme to transform back off the cut using the
transformation $y\to 1/y$ yielding HPLs with argument $-s/u$.

Now we have found the functions in all three analytic regions, and we can
transform the arguments to obtain all the different  permutations entering in
the two-loop amplitude~\eq{2loop}. For example $F(s,t)=F_3(s,t)$ since
we are in the region $s,t<0$, but $F(u,t)=F_1(u,t)$ since the first
argument is positive etc.

At this point, after summing all contributions, the two-loop
amplitude  will be a linear combination of harmonic polylogarithms
with different arguments. We therefore use the HPL programme again to
transform them all to the  same argument,  ensuring that we never
land on a branch cut in so doing. For example, for harmonic polylogarithms
of the form $H_{\vec{a},1}(x)$ (i.e.~where the defining string of numbers
ends in a `1')  we restrict ourselves to 
transformations of the form $y\rightarrow 1-y$ and $y\rightarrow
y/(y-1)$ which the HPL program performs assuming we are away from the
branch cut. 

Using the above techniques we obtain the following 
form for the two-loop finite remainder $ {\cal M}_4^{(2)} - \half ({\cal M}_4^{(1)})^2 $:
\begin{align} 
{\cal M}_4^{(2)} - \half ({\cal M}_4^{(1)})^2\ = &\  
\left({\kappa^2 \alpha_\epsilon \over 4} \right)^2 \Big[ 
s^2 f^{(s)}(y)
  \, + \, t^2 f^{(t)}(y) +\, u^2 f^{(u)}(y)\Big] 
   \ ,\label{2loopfull}
\end{align}
where
\begin{align}
f^{(s)}(y)=&\frac{L^4}{3}-\frac{2}{3} \log (1-y) L^3-\log ^2(1-y)
L^2+\pi ^2 L^2+\frac{2}{3} \log ^3(1-y) L\nn \\
&-4 \pi ^2 \log (1-y) L+8 S_{1,2}(y) L-4 \pi ^2
   \text{Li}_2(y)+ 8 S_{1,3}(y)-8
 S_{2,2}(y)-\frac{7 \pi ^4}{30}\nn \\
&+i \Big[\frac{2 \pi  L^3}{3}+2 \pi  \log (1-y) L^2
-2 \pi  \log ^2(1-y) L+8 \pi  \text{Li}_2(y) L\nn\\
&\qquad +\frac{4 \pi ^3 L}{3}-8 \pi 
   \text{Li}_3(y)+8 \pi  S_{1,2}(y)\Big]\ ,
\\[15pt]
f^{(t)}(y)=&\frac{2}{3} \log (1-y) L^3+\log ^2(1-y) L^2+4
\text{Li}_2(y) L^2-\pi ^2 L^2-\frac{2}{3} \log ^3(1-y) L\nn\\
&+4 \pi ^2 \log (1-y) L-8 \text{Li}_3(y)
   L+4 \pi ^2 \text{Li}_2(y)+8 \text{Li}_4(y)
-8 S_{1,3}(y)+\frac{\pi ^4}{2}\nn\\
&+i \Big[\frac{2 \pi  L^3}{3}-2 \pi  \log (1-y) L^2+2 \pi  \log
^2(1-y) L\nn\\
&\qquad-\frac{4 \pi ^3 L}{3}-8
   \pi  S_{1,2}(y)+8 \pi  \zeta (3)\Big]\ ,\\[15pt]
f^{(u)}(y)=&\frac{1}{3} \log ^4(1-y)-\frac{2}{3} L \log ^3(1-y)+L^2
\log ^2(1-y)-\frac{2}{3} L^3 \log (1-y)-4 L^2 \text{Li}_2(y)\nn\\
&+8 L \text{Li}_3(y)-8 \text{Li}_4(y)-8 L S_{1,2}(y)+8 S_{2,2}(y)-\frac{\pi
   ^4}{2}+L^2 \pi ^2\nn\\
&
+i
   \Big[-\frac{2 \pi  L^3}{3}-2 \pi  \log (1-y) L^2+2 \pi  \log
     ^2(1-y) L-8 \pi  \text{Li}_2(y) L \nn\\
&+\frac{4 \pi ^3 L}{3}-\frac{4}{3} \pi  \log
   ^3(1-y)-\frac{8}{3} \pi ^3 \log (1-y)+8 \pi  \text{Li}_3(y)-8 \pi
   \zeta (3)\Big]\ ,
\end{align}
where $y=-s/t$ and $L:=\log(-y)$.

Since the amplitude is invariant under crossing symmetry (arbitrary
permutations of the momenta or equivalently arbitrary permutations of
$s,t,u$) we must have
\begin{align}
  f^{(u)}(y)=f^{(u)}(1/y)=f^{(s)}(1-y)=
  f^{(t)}(y/(y-1))\ ,
\end{align}
which one can indeed verify as long as  one takes suitable care over the analytic
continuation in the manner outlined above. 

Simplifying slightly $f^{(u)}(y)$ by writing it as $k(y)+k(1/y)$ we
obtain the form of the amplitude given in~\eq{finalmente}.

\section{Derivation of \eqref{one-loop-chris2}}\label{app-D}
In this Appendix we derive \eqref{one-loop-chris2} from \eqref{1lcg} in the linearised 
gravity approximation. 
Upon expanding the metric about flat space, $g_{\mu \nu} (x) = \eta_{\mu \nu} + \kappa h_{\mu \nu}(x)$,
one finds that \eqref{1lcg} is equal to
\beqa
\label{one-loop-chris}
&& \kappa^2 \oint_{\cC} \! dx^\mu dx^\nu \, \lan \Gamma^\alpha_{\mu \beta} (x) \Gamma^\beta_{\nu \alpha} (y) \ran
\\ \nonumber
&& = \ {1\over2} \oint_{\cC} \! dx^\mu dx^\nu \, \left[ - \partial_\alpha^x \partial_\beta^x \lan h_\mu^\alpha (x) h_\nu^\beta (y) \ran \, + \,
\Box_x \lan h_{\mu \beta}(x) h_{\nu}^\beta (y) \ran \right ]
\ .
\eeqa
To perform the calculation in \eqref{one-loop-chris}
we choose  the de Donder gauge, where the  propagator
in $D=4-2 \epsuv$ dimensions is given by \eqref{don}. 
Boundary terms can be dropped as the contour is a closed loop. Doing this, one easily finds that%
\footnote{In \cite{Modanese2},  terms such as those appearing on the right hand side of 
\eqref{one-loop-chris22} are referred to as ``ultra-local".}
\beqa
\nonumber
 \kappa^2 \oint_{\cC} \! dx^\mu dy^\nu \, \lan \Gamma^\alpha_{\mu \beta} (x) \Gamma^\beta_{\nu \alpha} (y) \ran
 & = &  c (D) \, \oint_{\cC}\! dx_\mu dy^\mu \, \Box_{x} \Delta (x - y )
 \\ 
 &=& c(D) \oint_{\cC}  \ dx_\mu dy^\mu \delta^{(D)} (x-y)
\label{one-loop-chris22}
 \ ,
  \eeqa
where $c(D)$ is a numerical constant, finite as $D\to 4$. This is the result quoted in 
\eqref{one-loop-chris2}.


\newpage

\begin{thebibliography}{99}

\small

\bibitem{abdk}
  C.~Anastasiou, Z.~Bern, L.~J.~Dixon and D.~A.~Kosower,
  {\it Planar amplitudes in maximally supersymmetric Yang-Mills theory,}
  Phys.\ Rev.\ Lett.\  {\bf 91} (2003) 251602,
  {\tt hep-th/0309040}.

\bibitem{bds}
  Z.~Bern, L.~J.~Dixon and V.~A.~Smirnov,
  {\it Iteration of planar amplitudes in maximally supersymmetric Yang-Mills
  theory at three loops and beyond,}
  Phys.\ Rev.\  D {\bf 72} (2005) 085001,
  {\tt hep-th/0505205}.




\bibitem{ir1}
  A.~H.~Mueller,
  {\it On The Asymptotic Behavior Of The Sudakov Form-Factor,}
  Phys.\ Rev.\  D {\bf 20} (1979) 2037.

\bibitem{ir2}
  J.~C.~Collins,
  {\it Algorithm To Compute Corrections To The Sudakov Form-Factor,}
  Phys.\ Rev.\  D {\bf 22} (1980) 1478.


\bibitem{ir3}
  A.~Sen,
 {\it Asymptotic Behavior Of The Sudakov Form-Factor In QCD,}
  Phys.\ Rev.\  D {\bf 24} (1981) 3281.

\bibitem{ir4}
  G.~P.~Korchemsky,
  {\it Double Logarithmic Asymptotics in QCD,}
  Phys.\ Lett.\  B {\bf 217} (1989) 330.

\bibitem{ir5}
  S.~Catani and L.~Trentadue,
  {\it Resummation Of The QCD Perturbative Series For Hard Processes,}
  Nucl.\ Phys.\  B {\bf 327} (1989) 323.

\bibitem{ir6}
  L.~Magnea and G.~Sterman,
  {\it Analytic continuation of the Sudakov form-factor in QCD,}
  Phys.\ Rev.\  D {\bf 42} (1990) 4222.

\bibitem{ir7}
  S.~Catani,
  {\it The singular behaviour of {QCD} amplitudes at two-loop order,}
  Phys.\ Lett.\  B {\bf 427} (1998) 161,
  {\tt hep-ph/9802439}.

\bibitem{ir8}
  G.~Sterman and M.~E.~Tejeda-Yeomans,
  {\it Multi-loop amplitudes and resummation,}
  Phys.\ Lett.\  B {\bf 552} (2003) 48,
  {\tt hep-ph/0210130}.





\bibitem{2l5pt}
  Z.~Bern, M.~Czakon, D.~A.~Kosower, R.~Roiban and V.~A.~Smirnov,
  {\it Two-loop iteration of five-point N = 4 super-Yang-Mills amplitudes,}
  Phys.\ Rev.\ Lett.\  {\bf 97} (2006) 181601,
  {\tt hep-th/0604074}.


\bibitem{seven}
  Z.~Bern, L.~J.~Dixon, D.~A.~Kosower, R.~Roiban, M.~Spradlin, C.~Vergu and A.~Volovich,
{\it The Two-Loop Six-Gluon MHV Amplitude in Maximally Supersymmetric Yang-Mills
  Theory,}
  {\tt 0803.1465 [hep-th]}.



\bibitem{am}
  L.~F.~Alday and J.~Maldacena,
  {\it Gluon scattering amplitudes at strong coupling,}
  JHEP {\bf 0706} (2007) 064,
  {\tt 0705.0303 [hep-th]}.

\bibitem{fernando}
  L.~F.~Alday,
{\it Lectures on Scattering Amplitudes via AdS/CFT,}
{\tt 0804.0951 [hep-th]}.

\bibitem{am2}
  L.~F.~Alday and J.~Maldacena,
{\it Comments on gluon scattering amplitudes via AdS/CFT,}
  JHEP {\bf 0711} (2007) 068,
  {\tt 0710.1060 [hep-th]}.

\bibitem{lipa}
J.~Bartels, L.~N.~Lipatov and A.~S.~Vera,
  {\it BFKL Pomeron, Reggeized gluons and Bern-Dixon-Smirnov amplitudes,}
  {\tt 0802.2065}.


\bibitem{dks}
  J.~M.~Drummond, G.~P.~Korchemsky and E.~Sokatchev,
  {\it Conformal properties of four-gluon planar amplitudes and Wilson loops,}
  Nucl.\ Phys.\  B {\bf 795} (2008) 385,
  {\tt 0707.0243 [hep-th]}.

\bibitem{bht}
  A.~Brandhuber, P.~Heslop and G.~Travaglini,
  {\it MHV Amplitudes in N=4 Super Yang-Mills and Wilson Loops,}
  Nucl.\ Phys.\  B {\bf 794} (2008) 231,
  {\tt 0707.1153 [hep-th]}.

\bibitem{dhks4}
  J.~M.~Drummond, J.~Henn, G.~P.~Korchemsky and E.~Sokatchev,
  {\it On planar gluon amplitudes/Wilson loops duality,}
  Nucl.\ Phys.\  B {\bf 795} (2008) 52,
  {\tt 0709.2368 [hep-th]}.

\bibitem{dhks5}
  J.~M.~Drummond, J.~Henn, G.~P.~Korchemsky and E.~Sokatchev,
{\it Conformal Ward identities for Wilson loops and a test of the duality with
  gluon amplitudes,}
{\tt 0712.1223 [hep-th]}.

\bibitem{dhksbum}
  J.~M.~Drummond, J.~Henn, G.~P.~Korchemsky and E.~Sokatchev,
  {\it The hexagon Wilson loop and the BDS ansatz for the six-gluon amplitude,}
{\tt 0712.4138 [hep-th]}.

\bibitem{dhks6}
  J.~M.~Drummond, J.~Henn, G.~P.~Korchemsky and E.~Sokatchev,
  {\it Hexagon Wilson loop = six-gluon MHV amplitude,}
{\tt 0803.1466 [hep-th].}



\bibitem{gatheral}
  J.~G.~M.~Gatheral,
{\it Exponentiation Of Eikonal Cross-Sections In Nonabelian Gauge Theories,}
  Phys.\ Lett.\  B {\bf 133} (1983) 90.

\bibitem{taylor}
  J.~Frenkel and J.~C.~Taylor,
{\it Nonabelian Eikonal Exponentiation,}
  Nucl.\ Phys.\  B {\bf 246} (1984) 231.





\bibitem{magic}
  J.~M.~Drummond, J.~Henn, V.~A.~Smirnov and E.~Sokatchev,
  {\it Magic identities for conformal four-point integrals,}
  JHEP {\bf 0701} (2007) 064,
  {\tt hep-th/0607160}.




\bibitem{hep-th/9802162}
  Z.~Bern, L.~J.~Dixon, D.~C.~Dunbar, M.~Perelstein and J.~S.~Rozowsky,
  {\it On the relationship between Yang-Mills theory and gravity and its
  implication for ultraviolet divergences,}
  Nucl.\ Phys.\  B {\bf 530} (1998) 401,
{\tt hep-th/9802162}.

\bibitem{bes}
  N.~Beisert, B.~Eden and M.~Staudacher,
 {\it Transcendentality and crossing,}
  J.\ Stat.\ Mech.\  {\bf 0701} (2007) P021,
  {\tt hep-th/0610251}.

\bibitem{bkk}
  B.~Basso, G.~P.~Korchemsky and J.~Kotanski,
  {\it Cusp anomalous dimension in maximally supersymmetric Yang-Mills theory at
  strong coupling,}
  Phys.\ Rev.\ Lett.\  {\bf 100} (2008) 091601
  {\tt 0708.3933 [hep-th]}.


\bibitem{living-zvi}
  Z.~Bern,
  {\it Perturbative quantum gravity and its relation to gauge theory,}
  Living Rev.\ Rel.\  {\bf 5} (2002) 5,
  {\tt gr-qc/0206071}.




\bibitem{chalmers}
  G.~Chalmers,
  {\it On the finiteness of N = 8 quantum supergravity,}
{\tt hep-th/0008162}.


\bibitem{zero}
  N.~E.~J.~Bjerrum-Bohr, D.~C.~Dunbar, H.~Ita, W.~B.~Perkins and K.~Risager,
 {\it The no-triangle hypothesis for N = 8 supergravity,}
  JHEP {\bf 0612} (2006) 072,
  {\tt hep-th/0610043}.




\bibitem{Green:2006gt}
  M.~B.~Green, J.~G.~Russo and P.~Vanhove,
  {\it Non-renormalisation conditions in type II string theory and maximal supergravity},
  {\tt hep-th/0610299}.


\bibitem{bsgf1}
  Z.~Bern, L.~J.~Dixon and R.~Roiban,
{\it Is N = 8 supergravity ultraviolet finite?,}
  Phys.\ Lett.\  B {\bf 644} (2007) 265,
  {\tt hep-th/0611086}.


\bibitem{Green:2006yu}
  M.~B.~Green, J.~G.~Russo and P.~Vanhove,
  {\it Ultraviolet properties of maximal supergravity,}
  {\tt hep-th/0611273}.


\bibitem{bsgf2}
  Z.~Bern, J.~J.~Carrasco, L.~J.~Dixon, H.~Johansson, D.~A.~Kosower and R.~Roiban,
 {\it Three-Loop Superfiniteness of N=8 Supergravity,}
  Phys.\ Rev.\ Lett.\  {\bf 98} (2007) 161303,
  {\tt hep-th/0702112}.



\bibitem{bsgf3}
  Z.~Bern, J.~J.~Carrasco, D.~Forde, H.~Ita and H.~Johansson,
  {\it Unexpected Cancellations in Gravity Theories,}
  Phys.\ Rev.\  D {\bf 77} (2008) 025010,
 {\tt 0707.1035 [hep-th]}.





\bibitem{weinberg}
  S.~Weinberg,
  {\it Infrared photons and gravitons,}
  Phys.\ Rev.\  {\bf 140} (1965) B516.

\bibitem{blochnord}
  F.~Bloch and A.~Nordsieck,
  {\it Note on the Radiation Field of the electron,}
  Phys.\ Rev.\  {\bf 52} (1937) 54.

\bibitem{yfs}
  D.~R.~Yennie, S.~C.~Frautschi and H.~Suura,
 {\it The infrared divergence phenomena and high-energy processes,}
  Annals Phys.\  {\bf 13} (1961) 379.


\bibitem{klov}
  A.~V.~Kotikov, L.~N.~Lipatov, A.~I.~Onishchenko and V.~N.~Velizhanin,
  {\it Three-loop universal anomalous dimension of the Wilson operators in N =  4
  SUSY Yang-Mills model,}
  Phys.\ Lett.\  B {\bf 595} (2004) 521
  [Erratum-ibid.\  B {\bf 632} (2006) 754], 
  {\tt hep-th/0404092}.


\bibitem{ko}
  D.~N.~Kabat and M.~Ortiz,
  {\it Eikonal Quantum Gravity And Planckian Scattering,}
  Nucl.\ Phys.\  B {\bf 388} (1992) 570,
  {\tt hep-th/9203082}.


\bibitem{fpvv}
  M.~Fabbrichesi, R.~Pettorino, G.~Veneziano and G.~A.~Vilkovisky,
  {\it Planckian energy scattering and surface terms in the gravitational
  action,}
  Nucl.\ Phys.\  B {\bf 419} (1994) 147.






\bibitem{eik1}
  H.~Cheng and T.~T.~Wu,
  {\it High-energy elastic scattering in quantum electrodynamics,}
  Phys.\ Rev.\ Lett.\  {\bf 22} (1969) 666.

\bibitem{eik2}
  H.~D.~I.~Abarbanel and C.~Itzykson,
  {\it Relativistic eikonal expansion,}
  Phys.\ Rev.\ Lett.\  {\bf 23} (1969) 53.

\bibitem{eik3}
  M.~Levy and J.~Sucher,
  {\it Eikonal Approximation In Quantum Field Theory,}
  Phys.\ Rev.\  {\bf 186} (1969) 1656.





\bibitem{gsb}
  M.~B.~Green, J.~H.~Schwarz and L.~Brink,
 {\it N=4 Yang-Mills And N=8 Supergravity As Limits Of String Theories,}
  Nucl.\ Phys.\  B {\bf 198} (1982) 474.




\bibitem{dn}
  D.~C.~Dunbar and P.~S.~Norridge,
  {\it Calculation of graviton scattering amplitudes using string based methods,}
  Nucl.\ Phys.\  B {\bf 433} (1995) 181,
  {\tt hep-th/9408014}.



\bibitem{bk}
  Z.~Bern and D.~A.~Kosower,
  {\it The Computation of loop amplitudes in gauge theories,}
  Nucl.\ Phys.\  B {\bf 379} (1992) 451.



\bibitem{Bern:zx}
Z.~Bern, L.~J.~Dixon, D.~C.~Dunbar and D.~A.~Kosower, {\it One
Loop N Point Gauge Theory Amplitudes, Unitarity And Collinear
Limits,} Nucl.\ Phys.\ B {\bf 425} (1994) 217, {\tt
hep-ph/9403226}.



\bibitem{Bern:1994cg}
  Z.~Bern, L.~J.~Dixon, D.~C.~Dunbar and D.~A.~Kosower,
 {\it Fusing gauge theory tree amplitudes into loop amplitudes,}
  Nucl.\ Phys.\  B {\bf 435}, 59 (1995),
  {\tt hep-ph/9409265}.


\bibitem{bdpr}
  Z.~Bern, L.~J.~Dixon, M.~Perelstein and J.~S.~Rozowsky,
  {\it Multi-leg one-loop gravity amplitudes from gauge theory,}
  Nucl.\ Phys.\  B {\bf 546} (1999) 423,
  {\tt hep-th/9811140}.




\bibitem{nt}
  A.~Nasti and G.~Travaglini,
  {\it One-loop N=8 Supergravity Amplitudes from MHV Diagrams,}
  Class.\ Quant.\ Grav.\  {\bf 24}, 6071 (2007),
  {\tt 0706.0976 [hep-th]}.


\bibitem{bddk-selfdual}
  Z.~Bern, L.~J.~Dixon, D.~C.~Dunbar and D.~A.~Kosower,
 {\it One-loop self-dual and N = 4 superYang-Mills,}
  Phys.\ Lett.\  B {\bf 394} (1997) 105,
 {\tt hep-th/9611127}.





\bibitem{hep-ph/9905323}
  V.~A.~Smirnov,
  {\it Analytical result for dimensionally regularized massless on-shell  double
  box,}
  Phys.\ Lett.\  B {\bf 460} (1999) 397,
  {\tt hep-ph/9905323}.




\bibitem{hep-ph/9909506}
  J.~B.~Tausk,
  {\it Non-planar massless two-loop Feynman diagrams with four on-shell legs,}
  Phys.\ Lett.\  B {\bf 469} (1999) 225,
  {\tt hep-ph/9909506}.

%
\bibitem{Remiddi:1999ew}
  E.~Remiddi and J.~A.~M.~Vermaseren,
  {\it Harmonic polylogarithms,}
  Int.\ J.\ Mod.\ Phys.\  A {\bf 15} (2000) 725,
{\tt hep-ph/9905237}.




\bibitem{Green:2008kj}
  D.~Green,
  {\it Worldlines as Wilson Lines,}
 {\tt 0804.4450 [hep-th]}.


\bibitem{Modanese1}
  G.~Modanese,
  {\it Geodesic round trips by parallel transport in quantum gravity,}
  Phys.\ Rev.\  D {\bf 47} (1993) 502.

\bibitem{Modanese2}
  G.~Modanese,
  {\it Wilson loops in four-dimensional quantum gravity,}
  Phys.\ Rev.\  D {\bf 49} (1994) 6534,
  {\tt hep-th/9307148}.

\bibitem{mm1}
  Yu.~M.~Makeenko and A.~A.~Migdal,
  {\it Exact Equation For The Loop Average In Multicolor QCD,}
  Phys.\ Lett.\  B {\bf 88} (1979) 135
  [Erratum-ibid.\  B {\bf 89} (1980) 437].


\bibitem{mm2}
  Yu.~Makeenko and A.~A.~Migdal,
{\it Quantum Chromodynamics As Dynamics Of Loops,}
  Nucl.\ Phys.\  B {\bf 188} (1981) 269
  [Sov.\ J.\ Nucl.\ Phys.\  {\bf 32} (1980\ YAFIA,32,838-854.1980) 431.1980\ YAFIA,32,838].


\bibitem{dgo}
  N.~Drukker, D.~J.~Gross and H.~Ooguri,
  {\it Wilson loops and minimal surfaces,}
  Phys.\ Rev.\  D {\bf 60} (1999) 125006,
{\tt hep-th/9904191}.


\bibitem{mos}
  Y.~Makeenko, P.~Olesen and G.~W.~Semenoff,
 {\it Cusped SYM Wilson loop at two loops and beyond,}
  Nucl.\ Phys.\  B {\bf 748} (2006) 170,
 {\tt hep-th/0602100}.


\bibitem{polyakov}
  A.~M.~Polyakov,
{\it Gauge Fields As Rings Of Glue,}
  Nucl.\ Phys.\  B {\bf 164} (1980) 171.

\bibitem{cu1}
  I.~Y.~Arefeva,
  {\it Quantum Contour Field Equations,}
  Phys.\ Lett.\  B {\bf 93} (1980) 347.

\bibitem{cu2}
  S.~V.~Ivanov, G.~P.~Korchemsky and A.~V.~Radyushkin,
  {\it Infrared Asymptotics Of Perturbative QCD: Contour Gauges,}
  Yad.\ Fiz.\  {\bf 44}, 230 (1986)
  [Sov.\ J.\ Nucl.\ Phys.\  {\bf 44}, 145 (1986)].

\bibitem{cu3}
  V.~S.~Dotsenko and S.~N.~Vergeles,
  {\it Renormalizability Of Phase Factors In The Nonabelian Gauge Theory,}
  Nucl.\ Phys.\  B {\bf 169} (1980) 527.

\bibitem{cu4}
  R.~A.~Brandt, F.~Neri and M.~A.~Sato,
  {\it Renormalization Of Loop Functions For All Loops,}
  Phys.\ Rev.\  D {\bf 24} (1981) 879.

\bibitem{cu5}
  G.~P.~Korchemsky and A.~V.~Radyushkin,
{\it Loop Space Formalism And Renormalization Group For The Infrared Asymptotics
 Of QCD,}
  Phys.\ Lett.\  B {\bf 171}, 459 (1986).

\bibitem{cu6}
  I.~A.~Korchemskaya and G.~P.~Korchemsky,
  {\it On lightlike Wilson loops,}
  Phys.\ Lett.\  B {\bf 287} (1992) 169.

\bibitem{cu7}
  A.~Bassetto, I.~A.~Korchemskaya, G.~P.~Korchemsky and G.~Nardelli,
  {\it Gauge invariance and anomalous dimensions of a light cone Wilson loop in
  lightlike axial gauge,}
  Nucl.\ Phys.\  B {\bf 408} (1993) 62,
{\tt hep-ph/9303314}.











\bibitem{ftt}
  A.~Brandhuber, B.~Spence and G.~Travaglini,
  {\it From trees to loops and back,}
  JHEP {\bf 0601} (2006) 142,
  {\tt hep-th/0510253}.

%

\bibitem{Maitre:2005uu}
  D.~Maitre,
  {\it HPL, a Mathematica implementation of the harmonic polylogarithms,}
  Comput.\ Phys.\ Commun.\  {\bf 174} (2006) 222, 
{\tt hep-ph/0507152}.

\bibitem{Maitre:2007kp}
  D.~Maitre,
{\it Extension of HPL to complex arguments,}
{\tt hep-ph/0703052}.


\bibitem{Naculich:2008ew}
  S.~G.~Naculich, H.~Nastase and H.~J.~Schnitzer,
  {\it Two-loop graviton scattering relation and IR behavior in N=8
  supergravity,}
  {\tt 0805.2347 [hep-th]}.




\end{thebibliography}
\end{document}